\documentclass[12pt]{article}
\usepackage[margin=1in]{geometry}
\usepackage{amsmath, amssymb, amsbsy, amstext, amsthm, array, booktabs, IEEEtrantools, graphicx, cite, braket, float, tikz, subcaption, quantikz, setspace, listings, authblk, multirow}
\usepackage[
    colorlinks=true,
    citecolor=blue,
    urlcolor=blue,
    linkcolor=blue
    ]{hyperref}
\usetikzlibrary{decorations.pathreplacing}
\usepackage{grffile}[=v1]
\usepackage[aligntableaux=center]{ytableau}
\DeclareMathOperator{\trace}{tr}
\renewcommand{\braket}[2]{\langle #1 \vert #2 \rangle}
\newcommand{\ketbra}[2]{\vert #1 \rangle \langle #2 \vert}

\newcommand{\lbow}{\hspace{-3pt} \parbox{8pt}{\tikz{\draw[<-,line width=1pt] (0,0) -- (12pt,0)}} \hspace{-8pt} \left\{ \hspace{-4pt} \left\vert}
\newcommand{\rbow}{\right\vert \hspace{-4pt} \right\} \hspace{-13pt} \parbox{10pt}{\tikz{\draw[->,line width=1pt] (0,0) -- (12pt,0)}}}

\title{Perturbation theory, irrep truncations, and state preparation methods for quantum simulations of $SU(3)$ lattice gauge theory}

\author{Praveen Balaji}
\author{Cian\'an Conefrey-Shinozaki}
\author{Patrick Draper\thanks{pdraper@illinois.edu}}
\author{Jason K. Elhaderi}
\author{Drishti Gupta}
\author{Luis Hidalgo}
\author{Andrew Lytle}
\affil{Department of Physics, University of Illinois, Urbana, IL 61801}

\begin{document}

\maketitle

\begin{abstract}
    We study methods for efficient preparation of approximate ground states of $SU(3)$ lattice gauge theory on quantum hardware. Working in a variant of the electric basis, we introduce a refinement of the irrep truncation based on the energy density of site singlets, which provides a finer gradation of simulation complexity. Using strong-coupling perturbation theory as a guide, we develop simple ansatz circuits for ground state preparation and test them via classical simulation on small lattices, including the $2\times 2$ plaquette lattice in $d=2$ and the cube in $d=3$. We contrast state fidelities and resource requirements of variational methods against adiabatic state preparation and introduce a method that hybridizes the two approaches. Finally, we report on the public release of \texttt{ymcirc}---a package of tools for building $SU(3)$ circuits and processing measurements---and \texttt{pyclebsch}, a package for efficiently computing $SU(N)$ Clebsch-Gordan coefficients.
\end{abstract}

\tableofcontents

\section{Introduction}

A sufficiently advanced quantum computer could function as a unique and valuable tool for simulating quantum field theories (QFTs), providing access to nonperturbative phenomena in real time \cite{BauerEtal:2023:QS4HEP, BauerEtal:2023:QS_fund_particles_forces, DiMeglioEtal:2024:QC4HEP, HalimehEtal:2025:QS_out_of_equilibrium}. In particular, quantum simulations of lattice quantum chromodynamics (QCD) may be used to study parton distribution functions \cite{LammLawrenceYamauchi:2020:partonQC, ChenChenMeher:2025:PDF_on_QC}, hadronization \cite{FarrellIllaSavage:2025:hadronization, BarataRico:2025:jets_entropy_production}, and classically inaccessible parts of the QCD phase diagram \cite{DavoidiMuellerPowers:2023:QCphase_diagrams, ThanEtal:2024:1dQCD_phase_diagram}, all of which are directly relevant for natural and laboratory phenomena. Schematically, these future quantum simulations will require encoding a truncation of the target Hilbert space onto the quantum computer, preparing an initial quantum state, time-evolving the quantum state according to an approximation of a target Hamiltonian, and measuring observables. Much effort has gone into developing this framework over the last decade, particularly focused on encoding and time evolution.

Several approaches have been taken to  encode and time-evolve lattice gauge theories (LGTs) on digital quantum hardware. The most common starting point is the Kogut-Susskind (KS) Hamiltonian \cite{KogutSusskind:1975:KS_hamiltonian, ZoharBurrello:2015:LGT4QS}. Approaches differ, in part, in how bosonic gauge link degrees of freedom are truncated and encoded while addressing the Gauss law; a non-exhaustive list includes: working with eigenstates of the electric Hamiltonian (the electric basis), labeled by irreducible representations of a continuous gauge group \cite{ByrnesYamamoto:2006:sim_LGT_on_QC, CiavarellaKlcoSavage:2021:trailhead, BalajiEtal:2025:su3circuits, CiavarellaBauer:2024:large_Nc, CiavarellaBurbanoBauer:2025:efficient_large_Nc, RaychowdhuryStryker:2020:LSH_su2, KadamRaychowdhuryStryker:2023:LSH_su3_quarks, KadamEtal:2025:LSH_su3_trivalent, JiangKlcoMatteo:2025:cube_qudits, KavakiLewis:2024:square_to_triamond_su2, KavakiLewis:2025:false_vac_decay_triamond, KanEtal:2021:theta_term, KanNam:2022:QCD_QED_QC}; working with eigenstates of the magnetic Hamiltonian (the magnetic basis), labeled by group elements of a discrete gauge (sub)group \cite{LammLawrenceYamauchi:2019:general_methods, CarenaEtal:2022:improved_hamiltonians, GustafsonLammLovelace:2024:primitive_gates_su2bo, GustafsonEtal:2024:primitive_gates_su3S108}; and working with a hybrid of both bases \cite{HaaseEtal:2021:resource_efficient, DAndreaEtal:2024:new_basis_su2, GrabowskaKaneBauer:2025:fully_gauge_fixed_su2, BurbanoBauer:2024:LSH_on_graphs,k9p6-c649}. 

Although observables can  be calculated after time evolution on any state, a large class of interesting objects -- namely correlation functions in the ground state -- require the addressing the state preparation problem. 
Thus, quantum algorithms that prepare ground states of LGTs are necessary. 
A full framework for simulating QFTs (and in particular calculating scattering amplitudes in $\phi^4$ theory) was developed in \cite{JordanLeePreskill:2012:QA4QFT}. There, state preparation was possible because the ground state of the free theory was known and directly preparable. Then, the ground state of the interacting theory was built with adiabatic state preparation (ASP), where the free-theory ground state is evolved with a ``time-dependent'' Hamiltonian whose couplings are slowly turned on over time. In the case of $SU(N)$ LGT, a similar technique can be used; for example, the ground state at strong (weak) coupling is essentially the electric (magnetic) vacuum, and in principle the ground state at weaker (stronger) coupling can be prepared adiabatically from these starting points.

However, accurate ASP may require quantum circuits that are too deep for noisy intermediate-scale quantum (NISQ) computers. An alternative approach, the variational quantum eigensolver (VQE) \cite{PeruzzoEtal:2014:VQE_original, TillyEtal:2022:VQE_review}, was developed as a hybrid classical-quantum algorithm that can iteratively construct a reasonable ground state with a relatively shallower circuit. The algorithm takes a reference state (which for $SU(N)$ LGT can be, for example, the electric vacuum) that is evolved with a parameterized quantum circuit, called the ansatz. The energy of the output state is then measured, and a classical optimizer tunes the ansatz parameters; the process is iterated until convergence.  VQE has been used to probe a diverse range of phenomena in LGTs \cite{AtasEtal:2021:su2_hadrons_vqe, FarrellEtal:2023:preparations_1dQCD_axial, ZhangEtal:2023:vqe_superconducting, SchusterEtal:2024:schwinger_chemical_potential, DavoudiHsiehKadam:2024:scattering_wave_packets_hadrons, CrippaJansenRinaldi:2024:confinement_2dQED}, including $SU(3)$ LGT\cite{Ciavarella:2022:su3_state_prep}. It has also seen an assortment of enhancements related to optimal ansatz choice \cite{GrimsleyEtal:2019:ADAPT_VQE, FarrellEtal:2025:scattering_W_states}, scalability when increasing the lattice volume \cite{FarrellEtal:2024:scalable_schwinger_100qubits, FarrellEtal:2024:hadron_schwinger_112qubits, Zemlevskiy:2025:scalable_scalar_120qubits, ChernyshevEtal:2025:pathfinding_neutrinoless_2beta_decay}, and adjustments to the ansatz based on the lattice volume \cite{GustafsonEtal:2025:surrogate_scalable_vqe_schwinger}.

With an eye towards the future goal of simulating of lattice QCD, in this work we explore methods for  preparing ground states of $SU(3)$ LGT in two and three spatial dimensions. We use both VQE and ASP, and we develop two hybrid state preparation methods: one where the VQE reference state at a given coupling is the output state from a VQE performed at a higher coupling, and another where ASP is initialized with VQE. We use strong-coupling perturbation theory (PT) to motivate our VQE ans{\"a}tze. Although strong-coupling PT becomes unreliable at weak coupling, the target coupling for future simulations could be in a regime where PT is still a reasonable guide. We will explore this idea in some detail on small lattices and  benchmark ground state energies and eigenstate fidelities against classical exact diagonalization (ED) of the KS Hamiltonian.

Throughout this work we utilize the reduced electric basis of the KS Hamiltonian, described in \cite{BalajiEtal:2025:su3circuits}. This basis is an elaboration of the local multiplet basis of \cite{Banuls:2017ena,Klco:2019evd,CiavarellaKlcoSavage:2021:trailhead}, which leverages the gauge invariance of physical states to greatly reduce the original electric basis of~\cite{ByrnesYamamoto:2006:sim_LGT_on_QC}. Link degrees of freedom are described by $SU(3)$ irreducible representations (irreps), and lattice sites carry degrees of freedom that describe the gauge-singlet fusion of the irreps meeting at the site. One issue we explore is how state preparation accuracy and cost are affected by a chosen irrep truncation. A simple truncation is to cut on the electric energy of links. Roughly this energy grows with irrep dimensionality, so a variant of this truncation is to cut on the number of tensor indices in the irrep. We label $T_r$ the truncation that keeps up to all $r$-index tensor irreps. However, this truncation is extremely coarse, and small changes result in rapid variations in both  the maximum accessible energy density and the Hilbert space dimensionality. To improve on it we introduce a bound $B$ on the maximum allowed electric energy density at each site. This is a finer-grained truncation, and it leads to slower growth of the number of magnetic Hamiltonian matrix elements as the truncation is relaxed. Along the way, we describe improvements to the classical precomputation of the $SU(3)$ Clebsch-Gordan coefficients (CGCs), which are inputs to the classically precomputed magnetic matrix elements, over our previous implementation~\cite{BalajiEtal:2025:su3circuits}. The new implementation ``diagonalizes" repeated irreps in the direct sum decomposition into irreps of the symmetric group acting on repeated irreps in the direct product; this has the effect of reducing the number of nonzero magnetic matrix elements.  We have made the code that calculates these CGCs, \texttt{pyclebsch}, a public Python package.

Finally, we report briefly on the public release of \texttt{ymcirc}, a  Qiskit-based Python package of tools to build Trotterized time evolution circuits for $SU(3)$ LGT. The public version of the code supports circuit generation for the line-of-plaquettes (``d=3/2") lattice at the $T_{1}$ and $T_{2}$ irrep truncation, and for two-dimensional lattices at the $T_{1}$ irrep truncation, all with depth optimizations detailed in \cite{BalajiEtal:2025:su3circuits}. For state preparation, \texttt{ymcirc} also prepares paramterized versions of these evolution circuits for use in VQE.

\section{A softer irrep truncation}\label{sec:B-cutoff}

Below we introduce a truncation of the Hilbert space based on an energy associated with site-singlet  degrees of freedom, which feature in the reduced electric basis treatment of the Kogut-Susskind theory. We begin with a review of the reduced electric basis~\cite{BalajiEtal:2025:su3circuits}, which is based on the the local multiplet basis introduced to quantum simulations of gauge theories in~\cite{Banuls:2017ena,Klco:2019evd,CiavarellaKlcoSavage:2021:trailhead}.

\subsection{Review of the reduced electric basis}

We work with the pure-gauge Kogut-Susskind Hamiltonian,
\begin{equation}
    H = \sum_{\vec{s}} \left[ \frac{g^2}{2a} \sum_{i} E^2(\vec{s},\vec{e}_i) + \frac{1}{g^2a} \sum_{i<j} 2N - \Box(\vec{s},\vec{e}_i,\vec{e}_j) - \Box^\dag(\vec{s},\vec{e}_i,\vec{e}_j) \right]\label{KS_Hamiltonian}
\end{equation}
which is a suitable lattice discretization of the continuum Yang-Mills Hamiltonian. We work with cubic spatial lattices in various dimensions and set the spacing $a=1$.
 $E^2 \equiv E^b E^b$ ($b=1,\dots,N^2-1$) is the quadratic Casimir operator that acts on a lattice links, and
\begin{equation}
    \Box(\vec{s}, \vec{e}_i, \vec{e}_j) = \trace \left( U(\vec{s}, \vec{e}_i) U(\vec{s}+\vec{e}_i, \vec{e}_j) U^\dag(\vec{s} + \vec{e}_j, \vec{e}_i) U^\dag(\vec{s}, \vec{e}_j) \right)
\end{equation}
is the plaquette operator acting on four links that trace a minimal square. $\vec{s}$ gives the coordinates of a lattice site and $\vec{e}_i$ ($i=1,2,3$) are lattice unit vectors.

One can think of the Hilbert space as built from a (projection of) a tensor product of link Hilbert spaces. Each link Hilbert space is naturally thought of as a particle on the group manifold. In this picture the particle coordinates correspond to the magnetic basis. We work instead in the eigenbasis of $E^2$, where eigenstates $\ket{(R,r)}$ are labeled by an irreducible representation (irrep) $R$ of $SU(N)$ and a basis state $r$ of $R$.  Since the group admits left and right actions, a total link basis state is $\ket{(R,r_{\cal L})} \otimes \ket{(R,r_{\cal R})} \equiv \ket{(R,r_{\cal L},r_{\cal R})}$, which grants the ``left" and ``right" half-links independent degrees of freedom. Note that the $E^2$ eigenvalue, $E_{R_\ell}$, of a link state only depends on the irrep $R$. Henceforth, we will refer to this basis as the electric basis and $E_{R_\ell}$ as the electric energy of a link.

Gauss' law constrains the kinds of states that may appear on the lattice. We write physical lattice basis states in the electric basis as
\begin{equation}
    \hspace{-0.25in} \ket{\Lambda} = \bigotimes_{\vec{s}} \sum_{r} \left( \textstyle\prod_{\ell_{\cal L}(s)} \phi(r_{\cal L}) \right) \braket{\mathbf{1},\Gamma_s}{\otimes_{\ell_{\cal L}(s)} (\Bar{R}_\ell, \Tilde{r}_{\cal L}) \otimes_{\ell_{\cal R}(s)} (R_\ell,r_{\cal R}) } \ket{\otimes_{\ell_{{\cal L},{\cal R}}(s)} (R_\ell,r_{{\cal L},{\cal R}})}
    \label{physicalstate}
\end{equation}
where $\ell_{\cal L}(s)$ and $\ell_{\cal R}(s)$ denote all left and right half-links appearing at site $\vec{s}$, and $\ell_{{\cal L},{\cal R}}(s) = \ell_\mathcal{L}(s) \cup \ell_\mathcal{R}(s)$ denotes all half-links meeting at site $\vec{s}$. $\Bar{R}$ is a complex conjugate representation of $R$. (For real irreps, $\Bar{R}$ is equivalent to $R$.) $\Tilde{r}$ is a basis state of $\Bar{R}$ mapped from the basis state $r$ of $R$. There are conventional phases in this map between bases. These conventions are partially fixed by requiring the Clebsch-Gordan coefficients (CGCs) to be real, leaving only sign conventions to fix. These signs can be identified via the Clebsch-Gordan coefficient (CGC) $\braket{\mathbf{1}}{(R,r) \otimes (\Bar{R},\Tilde{r})}$, and $\phi(r)=\pm 1$ appearing in~(\ref{physicalstate}) is defined as the sign of this CGC. (We use \textbf{1} to denote the trivial representation.) 

The CGC $\braket{\mathbf{1},\Gamma_s}{\otimes_{\ell_{\cal L}(s)} (\Bar{R}_\ell, \Tilde{r}_{\cal L}) \otimes_{\ell_{\cal R}(s)} (R_\ell,r_{\cal R}) }$ appearing in~(\ref{physicalstate}) ensures each site forms a singlet in the electric basis, thus respecting Gauss' law. The particular singlet is determined by both the irreps meeting at the site and a multiplicity index $\Gamma$. The multiplicity index is essential because generically the group admits many invariant tensors of a given rank. For example, there are three unique singlets in $6\otimes\bar 6\otimes 6\otimes \bar 6$, and two in $8 \otimes 8 \otimes 8$. Note that all irrep basis states are summed over: therefore to  specify lattice basis states $\ket{\Lambda}$, it is sufficient to assigning irreps to each link, multiplicity indices to each site, and an $F$-order  to the singlet CGCs. (The $F$-order specifies the order of irreps in the tensor product, a nuisance which is described further below.) This specification of basis states is called the reduced electric basis. It relaxes the costly problem of assigning singlet-admitting irreps on each link along with a consistent multiplicity index, but it leaves room for irreps that fail to admit singlets and multiplicity indices higher than those allowed by the irreps meeting at a site. Thus the total Hilbert space reflects a balance between locality and gauge-invariance.

In any electric basis, the difficult part of the Hamiltonian is the magnetic term. 
We write the plaquette operator in the fundamental representation $f$, with basis states $\sigma$. The links and sites of the plaquette are assigned irreps and multiplicity indices $R_{\ell_k}$ and $\Gamma_{s_k}$, respectively, for $k=1,2,3,4$. The external links, or ``control" links~\cite{CiavarellaKlcoSavage:2021:trailhead} of the plaquette, meeting the plaquette at site $s_k$, have irreps denoted $\vec{C}_{s_k}$. Then the  plaquette operator matrix elements can be classically and efficiently pre-computed in the reduced electric basis from the formula~\cite{BalajiEtal:2025:su3circuits}
\begin{IEEEeqnarray*}{rCl}
    \hspace{-0.5in} \bra{\Lambda_n} \Box \ket{\Lambda_m} = \left( \prod_{k=1}^{4} \sqrt{\frac{\dim(R^m_{\ell_k})}{\dim(R^n_{\ell_k})}} \right) & & \lbow \begin{matrix} \bar{R}_{\ell_4}^m & \Gamma_{s_1}^m & \bar{R}_{\ell_1}^m \\ f & \vec{C}_{s_1} & \bar{f} \\ \bar{R}_{\ell_4}^n & \Gamma_{s_1}^n & \bar{R}_{\ell_1}^n \end{matrix} \rbow \quad \lbow \begin{matrix} R_{\ell_1}^m & \Gamma_{s_2}^m & \bar{R}_{\ell_2}^m \\ f & \vec{C}_{s_2} & \bar{f} \\ R_{\ell_1}^n & \Gamma_{s_2}^n & \bar{R}_{\ell_2}^n \end{matrix} \rbow \\
    \times & \ & \lbow \begin{matrix} R_{\ell_2}^m & \Gamma_{s_3}^m & R_{\ell_3}^m \\ f & \vec{C}_{s_3} & \bar{f} \\ R_{\ell_2}^n & \Gamma_{s_3}^n & R_{\ell_3}^n \end{matrix} \rbow \quad \lbow \begin{matrix} \bar{R}_{\ell_3}^m & \Gamma_{s_4}^m & R_{\ell_4}^m \\ f & \vec{C}_{s_4} & \bar{f} \\ \bar{R}_{\ell_3}^n & \Gamma_{s_4}^n & R_{\ell_4}^n \end{matrix} \rbow \yesnumber
    \label{eq:master_formula}
\end{IEEEeqnarray*}
where
\begin{IEEEeqnarray*}{rCl}
    \lbow \begin{matrix} R & G & S \\ f & \vec{C} & \bar{f} \\ R' & G' & S' \end{matrix} \rbow \ & = & \sum_{\sigma=1}^{\dim(f)} \sum_{r=1}^{\dim(R)} \sum_{r'=1}^{\dim(R')} \sum_{s=1}^{\dim(S)} \sum_{s'=1}^{\dim(S')} \sum_{\vec{c} \in \vec{C}} \phi(\sigma) \\
    & \times & \braket{R',r'}{(R,r) \otimes (f,\sigma)} \ \braket{S',s'}{(S,s) \otimes (\bar{f},\tilde{\sigma})} \\
    & \times & \braket{\mathbf{1},G}{(R,r) \otimes (S,s) \otimes (\vec{C},\vec{c})}_F \\
    & \times & \braket{\mathbf{1},G'}{(R',r') \otimes (S',s') \otimes (\vec{C},\vec{c})}_F \yesnumber
\end{IEEEeqnarray*}
are site factors, and the formula assumes a CGC phase convention such that
\begin{equation}
    \braket{(A,a) \otimes (f,\sigma)}{(B,b)} = \phi(a) \phi(\sigma) \phi(b) \braket{(\Bar{A},\Tilde{a}) \otimes (\Bar{f},\Tilde{\sigma})}{(\Bar{B}, \Tilde{b})}.
\end{equation}
The subscript $F$ found on the singlet CGCs reminds us that each irrep shown ($R$,$S$,$\vec{C}$) must be associated with a specific link on the lattice to create well-defined wavefunctions $\ket{\Lambda}$. An $F$-order is a conventional ordering of the links that meet at a site; it can be used to unambiguously associate each singlet CGC irrep to a lattice link. Thus, the order of the irreps in the CGC are understood to follow that of the links in the $F$-order.

We have described all the tools necessary, in principle, to carry out $SU(3)$ lattice simulations. However, the lattice volume and the Hilbert spaces of each link are  infinite. We must impose boundaries on the lattice as well as a truncation of the link Hilbert space. The former can be done with open or periodic boundary conditions. The latter can be done in a few ways. A common approach is to cap the underlying quantum numbers of SU(3) irreps, $R=(p,q)$. This places an upper bound on the electric energy of any link. Each irrep corresponds to a $p+q$-index tensor, so a qualitatively similar truncation is to cut on the value of $p+q$. We denote this truncation by $T_r$, where $r$ is the maximum value of $p+q$. For example, $T_1$ includes the trivial, fundamental, and antifundamental representations, and $T_2$ includes up to the two-index tensors.

\subsection{Site-singlet truncations}

The presence of site singlet degrees of freedom invites an elaboration on the electric link energy truncation. We define the electric energy density (more precisely the Casimir) associated with site $\vec{s}$ as $\sum_{\ell(s)} C_2({R_\ell})$, and we define a new truncation by placing an upper bound $B$ on this quantity. $B$ serves as a cutoff on the local electric energy density. $B$ effectively truncates the link irreps, but it also truncates the possible singlets that appear at a site. This truncation scheme can greatly reduce simulation costs, and in some cases we will see that it does so without significantly sacrificing simulation accuracy. It also allows for a finer gradation in the truncation: The jump in the number of accessible states as $T_i$ increases is in general much greater than the increase in the number of states as $B$ increases.

\begin{figure}[ht!]
    \centering
    \begin{subfigure}[b]{0.8\textwidth}
        \centering
        \includegraphics[width=0.8\textwidth]{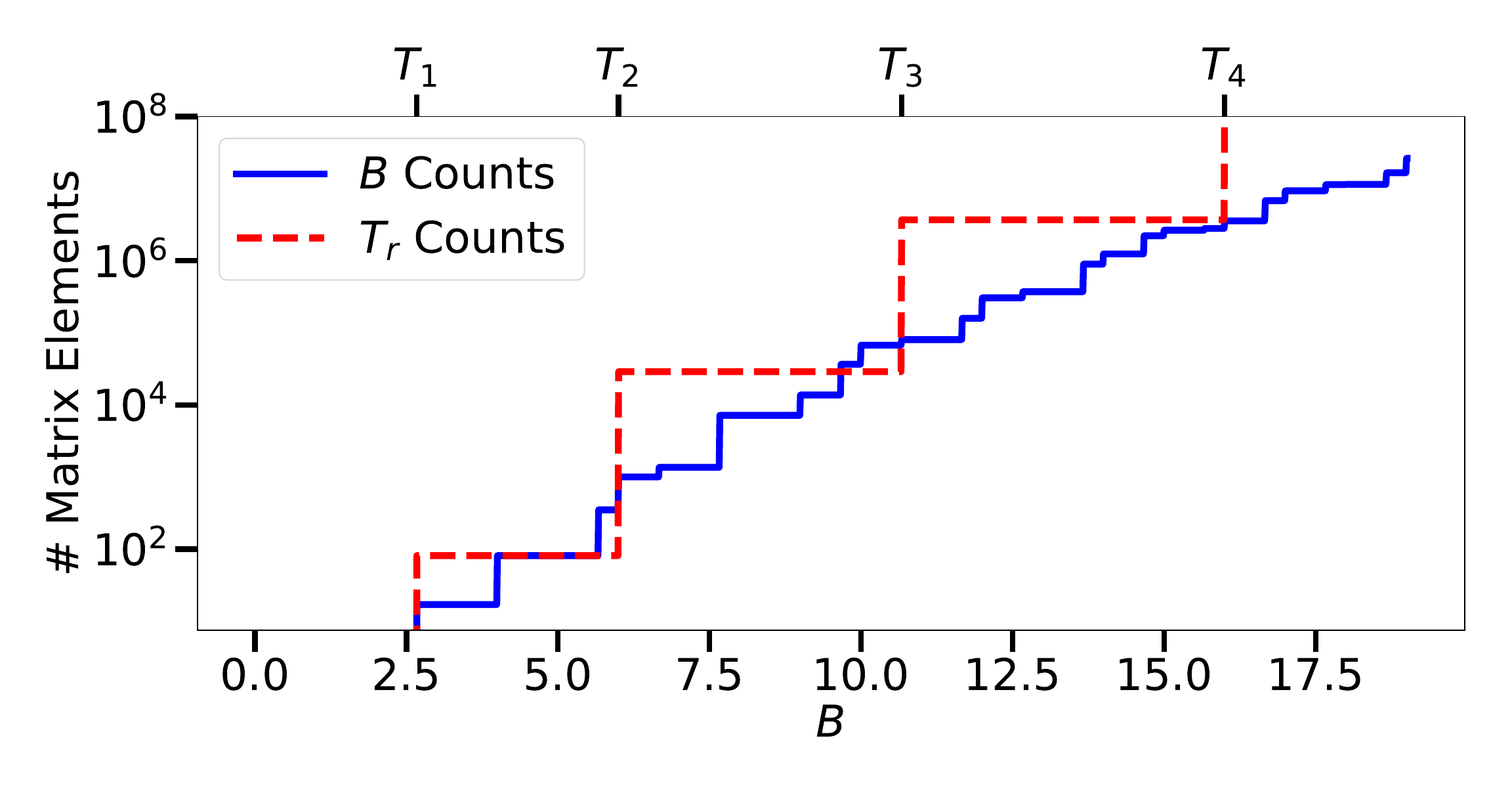}
    \end{subfigure}
    \begin{subfigure}[b]{0.8\textwidth}
        \centering
        \includegraphics[width=0.8\textwidth]{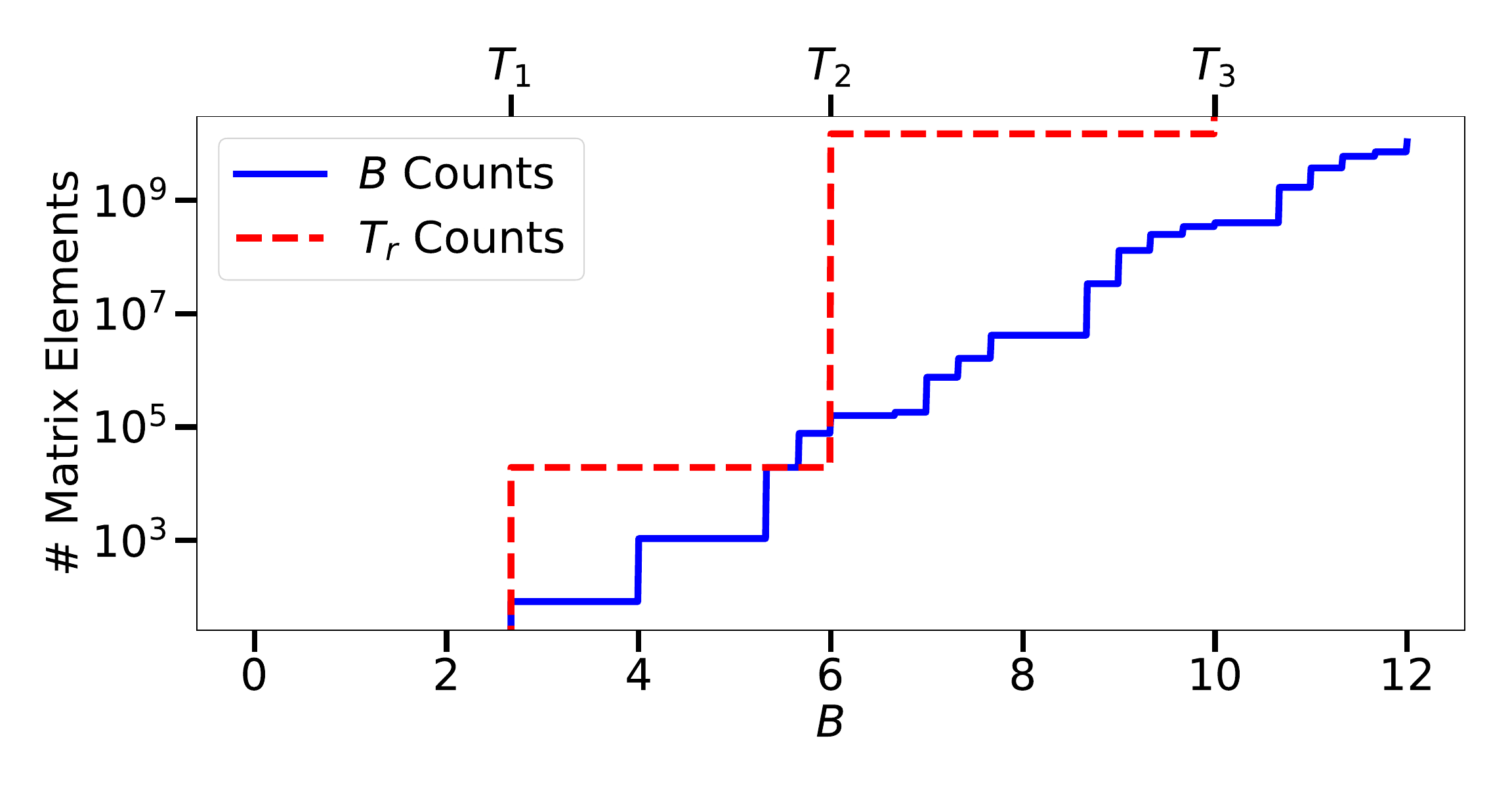}
    \end{subfigure}
    \begin{subfigure}[b]{0.8\textwidth}
        \centering
        \includegraphics[width=0.8\textwidth]{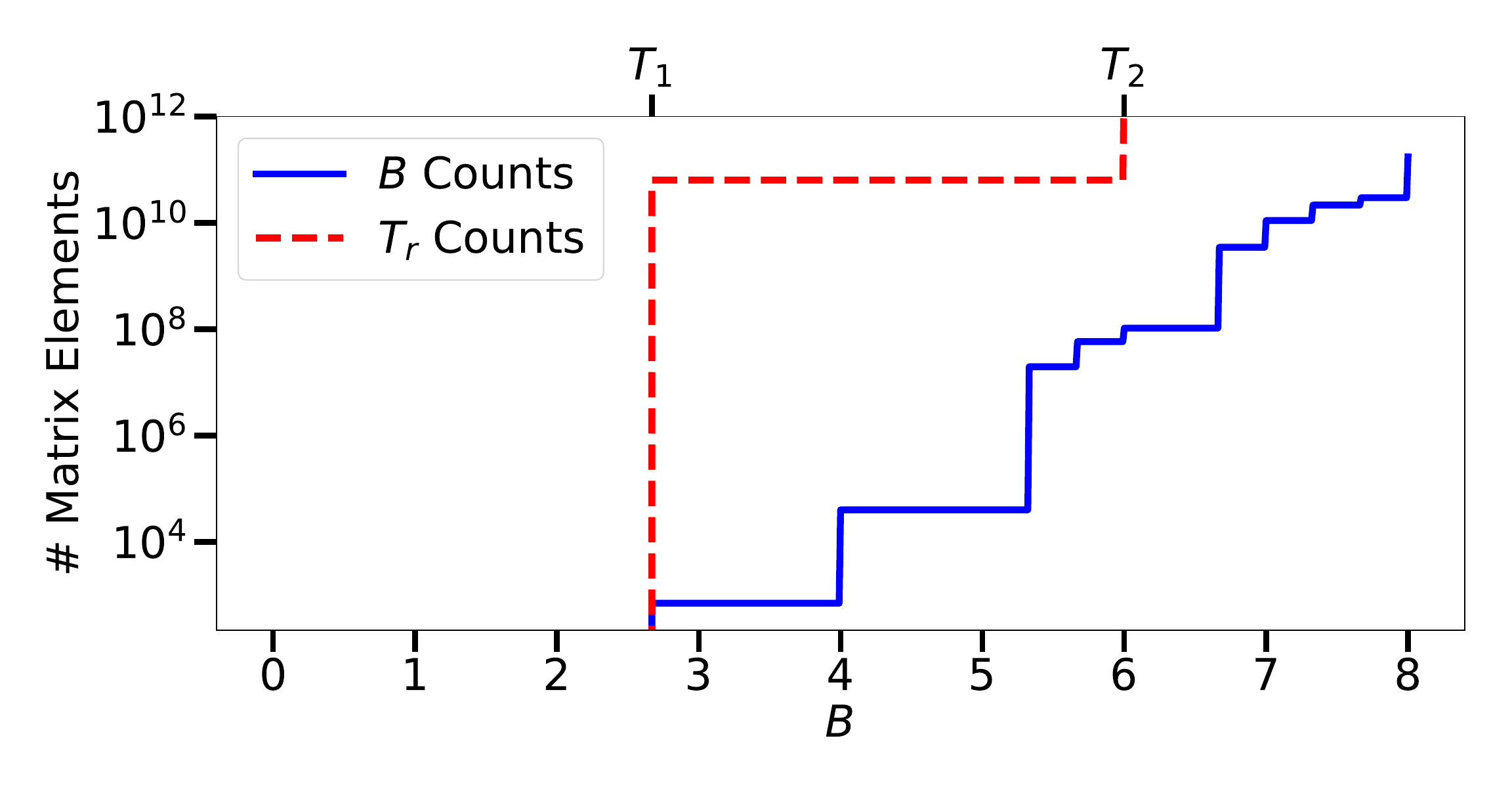}
    \end{subfigure}
    \caption{A comparison of $\Box$ matrix element counts in the $B$ truncation scheme versus the $T_r$ truncation for: $d=3/2$ (Top), $d=2$ (Middle), $d=3$ (Bottom). Points marking a specific $T_r$ indicate that $B$ has unlocked all $r$-index tensor irreps. Intervals where $B$ counts are greater than $T_r$ counts are those where $B$ has unlocked some but not all $r+1$-index tensor irreps.}
    \label{fig:unsym_counts}
\end{figure}
In Fig.~\ref{fig:unsym_counts} we compare $\Box$ matrix element counts in both Hilbert space truncation schemes for lattices of different dimensionalities. The $B$ truncations allow us to access irreps found in $T_r$ with orders-of-magnitude fewer matrix elements. In particular, by counting the number of gauge-invariant plaquette states, we estimate the $T_2$ truncation in $d=3$ to involve $\gtrsim 10^{20}$ matrix elements. However, with the $B$ truncation, {\emph{some}} states involving {\emph{all}} two-index tensor irreps can be accessed at $B > 6$ with $\sim 10^8$ matrix elements.

\begin{table}[ht]
    \centering
    {\renewcommand{\arraystretch}{1.2}
    \begin{tabular}[t]{c||c|c}
        $B$ & New Singlets & Multiplicity \\
        \hline\hline
        0 & $1 \otimes 1 \otimes 1$ & 1 \\
        \hline
        8/3 & $1 \otimes 3 \otimes \bar{3}$ & 1 \\
        \hline
        4 & $3 \otimes 3 \otimes 3 + \text{c.c.}$ & 1 \\
        \hline
        17/3 & $3 \otimes \bar{3} \otimes 8$ & 1 \\
        \hline
        6 & $1 \otimes 8 \otimes 8$ & 1 \\
        & $3 \otimes 3 \otimes \bar{6} + \text{c.c.}$ & 1 \\
        \hline
        20/3 & $1 \otimes 6 \otimes \bar{6}$ & 1 \\
        \hline
        23/3 & $3 \otimes 6 \otimes 8 + \text{c.c.}$ & 1 \\
        \hline
        9 & $8 \otimes 8 \otimes 8$ & 2 \\
        \hline
        29/3 & $6 \otimes \bar{6} \otimes 8$ & 1 \\
        & $3 \otimes 8 \otimes \bar{15} + \text{c.c.}$ & 1 \\
        \hline
        10 & $6 \otimes 6 \otimes 6 + \text{c.c.}$ & 1 \\
        & $3 \otimes \bar{6} \otimes 15 + \text{c.c.}$ & 1 \\
        \hline
        32/3 & $1 \otimes 15 \otimes \bar{15}$ & 1 \\
        & $3 \otimes 6 \otimes \bar{10} + \text{c.c.}$ & 1 \\
        \hline
        35/3 & $6 \otimes 8 \otimes 15 + \text{c.c.}$ & 1 \\
        \hline
        12 & $1 \otimes 10 \otimes \bar{10}$ & 1 \\
        & $8 \otimes 8 \otimes 10 + \text{c.c.}$ & 1 \\
        & $3 \otimes 15 \otimes 15 + \text{c.c.}$ & 1 \\
        & $6 \otimes 6 \otimes \bar{15} + \text{c.c.}$ & 1 \\
        \hline
        38/3 & $3 \otimes 10 \otimes \bar{15} + \text{c.c.}$ & 1 \\
        \hline
        41/3 & $8 \otimes 15 \otimes \bar{15}$ & 2 \\
        \hline
        14 & $6 \otimes \bar{15} \otimes \bar{15} + \text{c.c.}$ & 1 \\
        & $8 \otimes 8 \otimes 27$ & 1 \\
        \hline
        44/3 & $6 \otimes 10 \otimes 15 + \text{c.c.}$ & 1 \\
        & $6 \otimes \bar{10} \otimes 15 + \text{c.c.}$ & 1
    \end{tabular}
    \quad
    \begin{tabular}[t]{c||c|c}
        $B$ & New Singlets & Multiplicity \\
        \hline\hline
        44/3 & $3 \otimes \bar{15} \otimes 27 + \text{c.c.}$ & 1 \\
        & $6 \otimes 8 \otimes \bar{24} + \text{c.c.}$ & 1 \\
        & $6 \otimes \bar{6} \otimes 27$ & 1 \\
        \hline
        15 & $8 \otimes 10 \otimes \bar{10}$ & 1 \\
        & $3 \otimes 15 \otimes \bar{24} + \text{c.c.}$ & 1 \\
        \hline
        47/3 & $3 \otimes \bar{10} \otimes 24 + \text{c.c.}$ & 1 \\
        \hline
        16 & $1 \otimes 27 \otimes 27$ & 1 \\
        & $15 \otimes 15 \otimes 15 + \text{c.c.}$ & 2 \\
        & $6 \otimes 6 \otimes \bar{15}' + \text{c.c.}$ & 1 \\
        \hline
        50/3 & $10 \otimes 15 \otimes \bar{15} + \text{c.c.}$ & 1 \\
        & $6 \otimes 15 \otimes 27 + \text{c.c.}$ & 1 \\
        & $8 \otimes 15 \otimes 24 + \text{c.c.}$ & 1 \\
        & $1 \otimes 24 \otimes \bar{24}$ & 1 \\
        & $3 \otimes 10 \otimes \bar{15}' + \text{c.c.}$ & 1 \\
        \hline
        17 & $8 \otimes 10 \otimes 27 + \text{c.c.}$ & 1 \\
        & $6 \otimes \bar{15} \otimes 24 + \text{c.c.}$ & 1 \\
        \hline
        53/3 & $3 \otimes 24 \otimes 27 + \text{c.c.}$ & 1 \\
        & $6 \otimes 10 \otimes \bar{24} + \text{c.c.}$ & 1 \\
        & $8 \otimes 15 \otimes \bar{15}' + \text{c.c.}$ & 1 \\
        \hline
        18 & $10 \otimes 10 \otimes 10$ & 1 \\
        \hline
        56/3 & $1 \otimes 15' \otimes \bar{15}'$ & 1 \\
        & $15 \otimes \bar{15} \otimes 27$ & 2 \\
        \hline
        19 & $8 \otimes 27 \otimes 27$ & 2 \\
        & $3 \otimes 15' \otimes \bar{24} + \text{c.c.}$ & 1 \\
        & $15 \otimes 15 \otimes \bar{24} + \text{c.c.}$ & 2
    \end{tabular}}
    \caption{Singlets unlocked at several $d=3/2$ $B$ truncations, corresponding to those shown in the top panel of Fig.~\ref{fig:unsym_counts}.}
    \label{tab:3_2d_B_singlets}
\end{table}

\begin{table}[ht]
    \centering
    {\renewcommand{\arraystretch}{1.2}
    \begin{tabular}{c||c|c}
        $B$ & New Singlets & Multiplicity \\
        \hline\hline
        16/3 & $3 \otimes 3 \otimes \bar{3} \otimes \bar{3}$ & 2 \\
        \hline
        7 & $3 \otimes 3 \otimes 3 \otimes 8 + \text{c.c.}$ & 2 \\
        \hline
        22/3 & $3 \otimes \bar{3} \otimes \bar{3} \otimes \bar{6} + \text{c.c.}$ & 1 \\
        \hline
        26/3 & $3 \otimes \bar{3} \otimes 8 \otimes 8$ & 3 \\
        \hline
        9 & $3 \otimes 3 \otimes \bar{6} \otimes 8 + \text{c.c.}$ & 2 \\
        \hline
        28/3 & $3 \otimes 3 \otimes 6 \otimes 6 + \text{c.c.}$ & 1 \\
        & $3 \otimes \bar{3} \otimes 6 \otimes \bar{6}$ & 2 \\
        & $3 \otimes \bar{3} \otimes \bar{3} \otimes 15 + \text{c.c.}$ & 1 \\
        \hline
        10 & $3 \otimes 3 \otimes 3 \otimes \bar{10} + \text{c.c.}$ & 1 \\
        \hline
        32/3 & $3 \otimes 6 \otimes 8 \otimes 8 + \text{c.c.}$ & 3 \\
        \hline
        11 & $3 \otimes \bar{6} \otimes \bar{6} \otimes 8 + \text{c.c.}$ & 2 \\
        & $3 \otimes 3 \otimes 8 \otimes 15 + \text{c.c.}$ & 2 \\
        \hline
        34/3 & $3 \otimes 6 \otimes 6 \otimes \bar{6} + \text{c.c.}$ & 1 \\
        & $3 \otimes \bar{3} \otimes 6 \otimes 15 + \text{c.c.}$ & 1 \\
        & $\bar{3} \otimes \bar{3} \otimes \bar{6} \otimes 15 + \text{c.c.}$ & 2 \\
        \hline
        35/3 & $3 \otimes \bar{3} \otimes 8 \otimes 10 + \text{c.c.}$ & 1 \\
        \hline
        12 & $8 \otimes 8 \otimes 8 \otimes 8$ & 8 \\
        & $3 \otimes 3 \otimes \bar{6} \otimes 10 + \text{c.c.}$ & 1
    \end{tabular}}
    \caption{Singlets unlocked at several $d=2$ $B$ truncations, corresponding to those shown in the middle panel of Fig.~\ref{fig:unsym_counts}.}
    \label{tab:2d_B_singlets}
\end{table}

\begin{table}[ht]
    \centering
    {\renewcommand{\arraystretch}{1.2}
    \begin{tabular}{c||c|c}
        $B$ & New Singlets & Multiplicity \\
        \hline\hline
        20/3 & $1 \otimes 3 \otimes 3 \otimes 3 \otimes 3 \otimes \bar{3} + \text{c.c.}$ & 3 \\
        \hline
        8 & $3 \otimes 3 \otimes 3 \otimes 3 \otimes 3 \otimes 3 + \text{c.c.}$ & 5 \\
        & $3 \otimes 3 \otimes 3 \otimes \bar{3} \otimes \bar{3} \otimes \bar{3}$ & 6
    \end{tabular}}
    \caption{Singlets unlocked at several $d=3$ $B$ truncations, corresponding to those shown in the bottom panel of Fig.~\ref{fig:unsym_counts}.}
    \label{tab:3d_B_singlets}
\end{table}

We provide tables for the singlets that are unlocked at several values of $B$ at different spatial dimensionalities. Tables~\ref{tab:3_2d_B_singlets}, \ref{tab:2d_B_singlets}, and \ref{tab:3d_B_singlets} show the singlets accessed in the top, middle, and bottom panels of Fig.~\ref{fig:unsym_counts}, respectively. $d=3/2$ singlets in Tab.~\ref{tab:3_2d_B_singlets} are available at the same $B$ value in $d=2$ and $d=3$ by tensoring the appropriate number of 1's. Likewise, $d=2$ singlets in Tab.~\ref{tab:2d_B_singlets} are available at the same $B$ value in $d=3$ by tensoring two additional 1's. If we focus on $d=2$, then at $B=16/3$, the $B$ irrep truncation coincides with $T_1$. Meanwhile, $B=6$ allows all two-index irreps in the truncation, but it is not equivalent to $T_2$. For instance, $8 \otimes 8 \otimes 8 \otimes 8$, which would appear in $T_2$, is still locked until $B=12$, and the highest energy $T_2$ singlet, $6 \otimes 6 \otimes \bar{6} \otimes \bar{6}$, would not appear until $B=40/3$.

Conceptually, the $B$ truncation should be regarded as a cutoff in the sense of effective field theory, since (normalized to the number of links per vertex) it controls local energy densities, defined as an average at the lattice scale. Defining the energy density by an average over multiple sites might also be of use. The truncation needed for a given large-lattice simulation to accurately compute certain continuum observables is an important question we do not address here; however, we will investigate convergence properties of ground state energies with $B$ at different $g$ in various small lattices.

\subsection{CGC improvements}
Pre-computing the plaquette matrix elements rely on efficient computation of $SU(3)$ CGCs. Our algorithm, described in~\cite{BalajiEtal:2025:su3circuits}, is based on the algorithm of~\cite{AlexEtal:2011:SUN_CGCs}. We have extended it as described in this subsection based on the algorithms of \cite{vanLeeuwenCoehnLisser:1992:LIE_manual} (adapted from \cite{Fonseca:2021:GroupMath}) as well as \cite{AlcockZeilingerWeigert:2017:young_symmetrizers}.  This improvement results in a further reduction in the number of nonzero matrix elements as the spatial dimension and $B$ cutoff are increased.

Our strategy for improving the CGC calculation is based on the observation that even after fixing various phase conventions, CGCs are not necessarily uniquely fixed.  Of particular interest is when the same $SU(N)$ irrep appears multiple times in a direct product. More precisely, let $R$ be an $SU(N)$ irrep, and consider the direct-sum decomposition $R^{\otimes n} = \oplus_i r_i$, where each $r_i$ is also an $SU(N)$ irrep. Then a basis can be chosen so that each $r_i$ transforms in an irrep $\lambda$ of the symmetric group $S_n$. The CGCs of the basis states of $r_i$ can then be symmetrized with (Hermitian) Young symmetrizers \cite{AlcockZeilingerWeigert:2017:young_symmetrizers}, and they will manifestly display the symmetries of the $S_n$ irreps $\lambda$. This procedure is described in greater detail in appendix \ref{appendix:improved cgcs}.

\begin{figure}[ht!]
    \centering
    \includegraphics[width=0.9\textwidth]{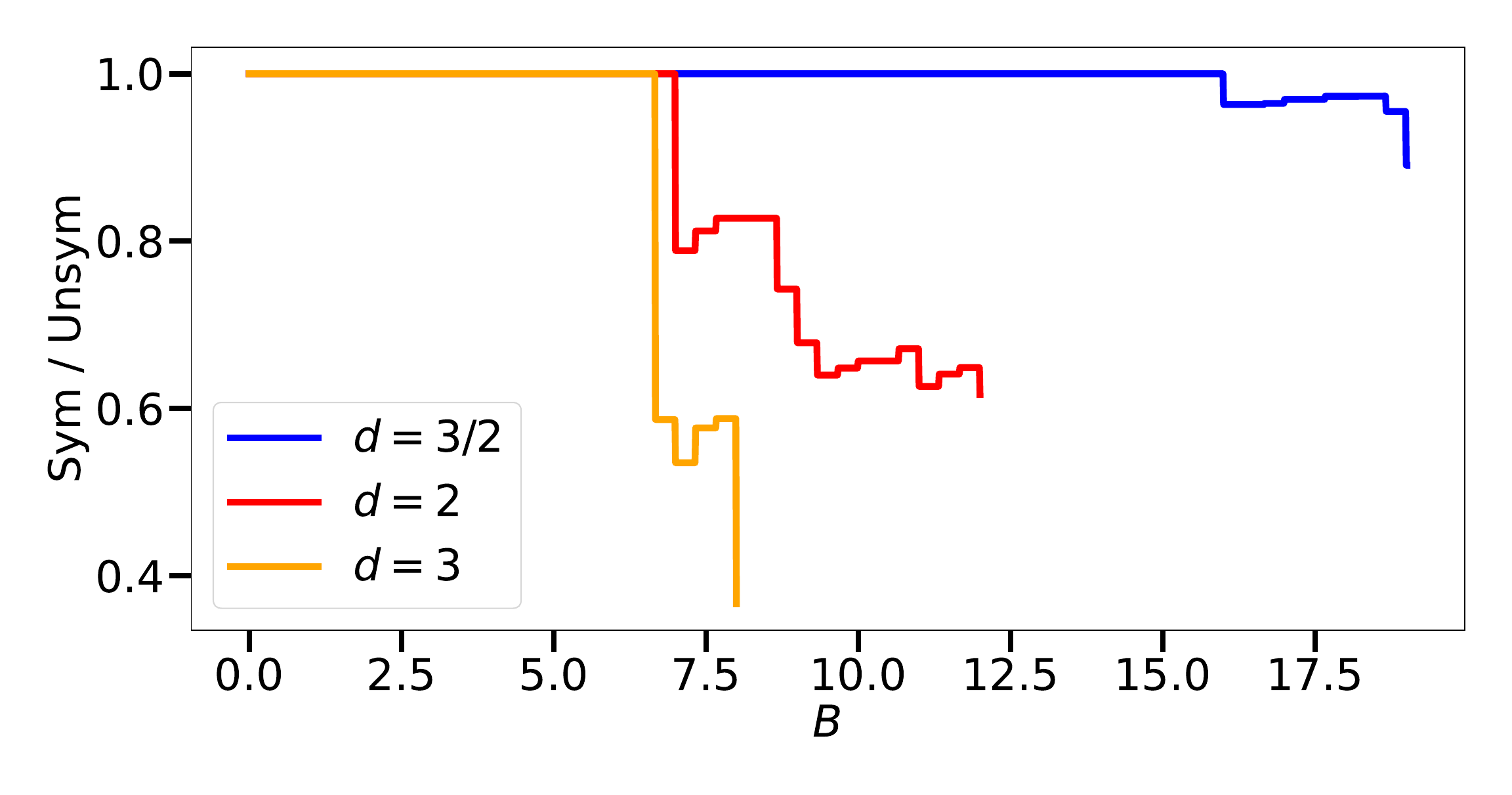}
    \caption{The reduction in $\Box$ matrix element counts  using symmetrized CGCs (``Sym'') versus  unsymmetrized CGCs (``Unsym''). Each curve shows the ratio $\frac{\text{Sym}}{\text{Unsym}}$ for different site-singlet energy cutoffs ($B$) in $d$ dimensions. Sym counts are never higher than Unsym counts, suggesting that symmetrized CGCs are more efficient for both matrix element calculations and quantum simulations.}
    \label{fig:sym_counts}
\end{figure}

We have implemented this ``diagonalization of irreps of permutations" in our CGC code, and in Fig.~\ref{fig:sym_counts} we compare the number of $\Box$ matrix elements found before and after CGC symmetrization for different $B$. We see that the symmetrized CGCs begin to reduce the matrix element counts when higher multiplicity indices begin to appear.

\clearpage
\section{Quantum circuits}\label{sec:quantum_circuits}

In this section we review our quantum circuit construction from \cite{BalajiEtal:2025:su3circuits}. There, we developed Trotterized time evolution circuits for $SU(3)$ LGT in two and three spatial dimensions. Our state preparation methods allow us to make use of these circuits, either directly or by adding additional parameters.

\subsection{Encoding gauge degrees of freedom}

Link irreps and site multiplicities are encoded with dense binary bit strings. For example, in a $d=3$, $B=8$ truncation simulation, the set of possible link irreps is $\{ 1,3,\bar{3},6,\bar{6},8 \}$. Site singlet multiplicities can be at most six-fold, and they can be indexed with the integers $\{ 0,1,2,3,4,5 \}$. The encoding of these data is shown in table~\ref{tab:example_encodings}.\footnote{The particular choice of map between bit strings and irreps or multiplicity indices is arbitrary. Our simulation code described in section \ref{sec:ymcirc} allows users to specify alternatives to the encodings given in table~\ref{tab:example_encodings}.} Thus, for this simulation, three qubits would be required for every link, and three qubits would be required for every site. For truncations where no singlet has a multiplicity, no qubits are added for sites.
\begin{table}[h]
    \centering
    \begin{tabular}{c|c}
        Irrep & Bit String \\
        \hline\hline
        1 & $\ket{000}$ \\
        3 & $\ket{001}$ \\
        $\bar{3}$ & $\ket{010}$ \\
        6 & $\ket{011}$ \\
        $\bar{6}$ & $\ket{100}$ \\
        8 & $\ket{101}$
    \end{tabular}
    \qquad
    \begin{tabular}{c|c}
        Multiplicity Index & Bit String \\
        \hline\hline
        0 & $\ket{000}$ \\
        1 & $\ket{001}$ \\
        2 & $\ket{010}$ \\
        3 & $\ket{011}$ \\
        4 & $\ket{100}$ \\
        5 & $\ket{101}$
    \end{tabular}
    \caption{Example link irrep and site multiplicity index encodings for a $d=3$, $B=8$ truncation simulation. Generally, we adopt a dense binary encoding of these reduced electric basis data.}
    \label{tab:example_encodings}
\end{table}

The KS Hamiltonian contains the operators $E^2(\vec{s},\vec{e}_i)$, $\Box(\vec{s},\vec{e}_i,\vec{e}_j)$, and the Hermitian conjugate $\Box^\dag(\vec{s},\vec{e}_i,\vec{e}_j)$. In the encoded reduced electric basis, $E^2$ remains a diagonal operator; namely, it is a sum containing operators of the form
\begin{equation}
    I \otimes \cdots \otimes I \otimes \ketbra{R_\ell}{R_\ell} \otimes I \otimes \cdots \otimes I 
\end{equation}
where $\ketbra{R_\ell}{R_\ell}$ projects onto an irrep $R$ of a link $\ell(\vec{s},\vec{e}_i)$. For instance, $\ketbra{001}{001}$ projects onto the irrep 3 in the example above. Meanwhile, $\Box$ is a non-diagonal operator that acts on a plaquette $P(\vec{s},\vec{e}_i,\vec{e}_j)$. The state of a plaquette includes the irreps on its four links, $\ell_1,\ell_2,\ell_3,\ell_4$, the irreps on its control links, $\vec{C}_1,\vec{C}_2,\vec{C}_3,\vec{C}_4$, and the multiplicity indices on its four sites, $s_1,s_2,s_3,s_4$. Hence, $\Box$ is a sum containing operators of the form\footnote{The example term given to illustrate the form of $\Box$ assumes a $d = 3$ lattice with periodic boundary conditions, where there are four control links connected to each site. For other lattice geometries, the number of control links per site may differ.}
\begin{multline}
    I \otimes \cdots \otimes I \otimes \\ \ket{\Gamma'_{s_1}} \dots \ket{\Gamma'_{s_4}} \ket{R'_{\ell_1}} \dots \ket{R'_{\ell_4}} \ket{R_{C_{1,1}}} \dots \ket{R_{C_{4,4}}} \bra{\Gamma_{s_1}} \dots \bra{\Gamma_{s_4}} \bra{R_{\ell_1}} \dots \bra{R_{\ell_4}} \bra{R_{C_{1,1}}} \dots \bra{R_{C_{4,4}}}\\\otimes I \otimes \cdots \otimes I
\end{multline}
which induce transition one plaquette state to another. (The corresponding term in $\Box^\dag$ induces the reverse transition.) The coefficients of $E^2$ are $E_{R_\ell}$, the quadratic Casimir eigenvalue for an irrep $R$ of $SU(3)$. The matrix elements of $\Box$ are calculated with Eq.~(\ref{eq:master_formula}).

Our circuits for time evolution and state preparation make use of the first-order Trotterized time evolution operator:
\begin{equation}
    \left( \prod_{j} e^{-i H_{E,j} \Delta t} \prod_{k} e^{-i H_{B,k} \Delta t} \right)^n
\end{equation}
Here, $H_{E,j}$ is a single link irrep projection operator, and $H_{B,k}$ is a single (Hermitian) plaquette state transition, both in the form of the operators mentioned above with the appropriate coefficients. For time evolution, or adiabatic state preparation with a time dependent coupling $g(t)$, $\Delta t$ is used as a time step parameter equal to $\frac{t}{n}$, and Trotter error scales as $\mathcal{O}(\Delta t^2)$. For our variational state preparation methods, $\Delta t$ may become a more general parameter $\theta$, such that each $H_{E,j}$ and $H_{B,k}$ may be multiplied by a parameter $\theta_j$ and $\theta_k$, respectively. Moreover, each Trotter step operator can have its own set of parameters.  As will be described in Sec.~\eqref{sec:em-emem-ansatz}, the number of parameters can be systematically controlled, with the $EM$ ans\"{a}tze requiring $2n$ unique parameters, and the multi-Givens ansatz using the same $H_{B,k}$ parameters for each plaquette.

\subsection{Electric evolution}

Each operator $H_{E,j}$ is proportional to a projector. To construct them, we start by defining the single-qubit projection operators
\begin{equation}
    \pi^0 = \ketbra{0}{0} = \frac{I+Z}{2}, \qquad \pi^1 = \ketbra{1}{1} = \frac{I-Z}{2}.
\end{equation}
Because the number of qubits required to encode link irreps is low, we opt for the Pauli operator definition of the projectors, $\frac{I \pm Z}{2}$, and build circuits with a Pauli string decomposition. For example, suppose
\begin{equation}
    H_{E,j} = \ketbra{001}{001} \theta = \pi^0 \pi^0 \pi^1 \theta,
\end{equation}
 with identity operators suppressed. This operator has Pauli string decomposition
 \begin{equation}
    \pi^0 \pi^0 \pi^1 \theta = \frac{\theta}{8}( III - IIZ - IZZ + IZI + ZZI - ZZZ - ZIZ + ZII )
 \end{equation}
(Notice that the Pauli strings have been written in a Gray code \cite{Nielsen_Chuang_2010} order to cancel eventually adjacent CX gates.) The $III$ string can be discarded as it generates a global phase. Each Pauli string commutes with each other, so the total matrix exponential can be written as
\begin{equation}
    e^{-i H_{E,j}} = e^{i\frac{\theta}{8}IIZ} \cdots e^{-i\frac{\theta}{8}ZII}
\end{equation}
without extra Trotter error. Exponentials whose Pauli string contains one $Z$ can be implemented as an $R_Z$ gate on the corresponding qubit. If the Pauli string contains more than one $Z$, then CX gates are required to transmit a parity sign flip onto an $R_Z$ gate, as shown in figure~\ref{fig:example_electric_gate}.
\begin{figure}[ht]
    \centering
    $e^{i \frac{\theta}{8} ZZZ}$ =
    \begin{quantikz}[align equals at=2, row sep={24pt,between origins}]
        \lstick{$q_2$} & \ctrl{2} & & & & \ctrl{2} & \\
        \lstick{$q_1$} & & \ctrl{1} & & \ctrl{1} & & \\
        \lstick{$q_0$} & \targ{} & \targ{} & \gate{R_Z(-\frac{\theta}{4})} & \targ{} & \targ{} &
    \end{quantikz}
    \caption{Example three-qubit gate used for electric evolution. The CX gates  calculate the parity of an ingoing bit string to apply the appropriate sign of the $R_Z$ rotation angle.}
    \label{fig:example_electric_gate}
\end{figure}
Thus, for each operator $e^{-i H_{E,j}}$ acting on $n$ qubits, there are $2^n-2$ CX gates.

\subsection{Magnetic evolution}
\label{sec:mag evol}

Each operator $H_{B,k}$ is a product of ladder and projection operators. We construct them by first defining single-qubit ladder operators
\begin{equation}
    \sigma^+ = \ketbra{1}{0} = \frac{X - iY}{2}, \qquad \sigma^- = \ketbra{0}{1} = \frac{X + iY}{2}.
\end{equation}
The number of qubits that are nontrivially acted on by $H_{B,k}$ is typically $\mathcal{O}(10)$ for the simulations reported on in this paper.  As a result, a Pauli string decomposition of $H_{B,k}$ would  lead to very deep magnetic evolution circuits. Instead, a diagonalization procedure can be used such that each operator $e^{-i H_{B,k}}$ amounts to a Givens rotation \cite{CiavarellaKlcoSavage:2021:trailhead}: a generalized $R_X$ gate that rotates two---and only two---bit strings made of the qubits that encode a plaquette's state. For example, consider the operator
\begin{equation}
    H_{B,k} = \ketbra{01011010}{00000000}\theta + \text{h.c.,}
\end{equation}
with identity operators suppressed. We identify this operator as belonging to the PLPLLPLP ``LP family,'' where ``P'' stands for projector and ``L'' stands for ladder~\cite{BalajiEtal:2025:su3circuits}. Generally, an LP family operator can be diagonalized by first choosing a ``pivot'' qubit $q'$ on any qubit that has an L. Then, apply CX gates controlled on $q'$, targeting each other qubit with an L, thus entangling a change on the pivot qubit with the rest of the qubits with an L. This product of CX gates forms the diagonalizing unitary $V$ such that $e^{-i H_{B,k}} = V e^{-i \tilde{H}_{B,k}} V^\dag$, and $\tilde{H}_{B,k}$ is a product of projectors everywhere except at the pivot qubit, now acted on by an $X = \sigma^+ + \sigma^-$ gate. For the running example, this yields $\tilde{H}_{B,k} = \pi^0 X \pi^0 \pi^0 \pi^0 \pi^0 \pi^0 \pi^0$ if $q'$ is chosen to be the first L, reading left to right. $e^{-i \tilde{H}_{B,k}}$ is a multi-controlled $R_X$ gate, and the complete circuit is shown in figure~\ref{fig:example_magnetic_gate}.
\begin{figure}[ht]
    \centering
    $e^{-i \theta (\pi^0 \sigma^+ \pi^0 \sigma^+ \sigma^+ \pi^0 \sigma^+ \pi^0 + \text{h.c.})} = $
    \begin{quantikz}[align equals at=4.5, row sep={24pt,between origins}]
        \lstick{$q_7$} & & & & \ocontrol{} & & & & \rstick{(P)} \\
        \lstick{$q_6$} & \targ{} & & & \ocontrol{} & & & \targ{} & \rstick{(L)} \\
        \lstick{$q_5$} & & & & \ocontrol{} & & & & \rstick{(P)} \\
        \lstick{$q_4$} & & \targ{} & & \ocontrol{} & & \targ{} & & \rstick{(L)} \\
        \lstick{$q_3$} & & & \targ{} & \ocontrol{} & \targ{} & & & \rstick{(L)} \\
        \lstick{$q_2$} & & & & \ocontrol{} & & & & \rstick{(P)} \\
        \lstick{$q_1$} & \ctrl{-5} & \ctrl{-3} & \ctrl{-2} & \gate{R_X(2\theta)} & \ctrl{-2} & \ctrl{-3} & \ctrl{-5} & \rstick{(L)} \\
        \lstick{$q_0$} & & & & \octrl{-7} & & & & \rstick{(P)}
    \end{quantikz}
    \caption{Example eight-qubit Givens rotation circuit used for magnetic evolution. This circuit performs a generalized $R_X$ rotation on only the bit strings $\ket{00000000}$ and $\ket{01011010}$.}
    \label{fig:example_magnetic_gate}
\end{figure}

Due to the number of possible plaquette state transitions found with Eq.~\ref{eq:master_formula}, the magnetic evolution is the most expensive part of our circuits to simulate. However, as detailed in \cite{BalajiEtal:2025:su3circuits}, we are able to implement several optimization strategies that reduce circuit size and depth. Among them is control pruning, where controls in the multi-controlled $R_X$ gates are eliminated via an iterative algorithm that identifies a smaller set of controls that still avoids rotating all other physical plaquette states (the number of which is much smaller than $2^n$). Another strategy is the use of ancillas to build Givens rotation in terms of single- and two-qubit gates via the v-chain method, which requires two less ancillas than the number of controls \cite{bennakhi-2024-analyzin-quantum-circuit, BarencoEtal:1995:elementary_gates}. Together, these strategies allow an $n$-qubit multi-controlled $R_X$ gate to be built with $6n-12$ CX gates, where $n$ can be minimized with control pruning.
\section{Ground state preparation: Methods}
\label{sec:methods}
Ground state preparation is a vital first step in exploring the dynamics of physically reasonable excited states. The goal, of course, is to  leverage quantum computation beyond feasible classical algorithms. 

However, in the near term quantum simulations will be limited to small lattice systems. In these cases the ground state can also be obtained by classical  exact diagonalization methods, as reviewed in Appendix \ref{appendix:exact_diagonalization}. This provides a valuable benchmark for the performance of different quantum algorithms and truncation methods. 

We will use perturbation theory as a guide in developing efficient quantum circuits for approximate ground state preparation. Since we are working in a variant of the electric basis, the relevant perturbative expansion is the strong-coupling expansion in  small $\frac1g$, which we now review.

\subsection{Strong-coupling perturbation theory}
\label{sec:PT}
The KS Hamiltonian in Eq.~(\ \ref{KS_Hamiltonian}) can be rewritten  as $\hat H = \hat H_0 + \lambda \hat V$, where
\begin{align}
      \hat H_0 &= \frac{g^2}{2a^{d-2}} \sum_\text{links}  \hat  E_\ell ^2,\\
    \lambda, &= -\frac{1}{a^{4-d}g^2}\\
     \hat  V &= \sum_\text{plaq}\hat \square_p + \hat \square_p^\dagger. 
\end{align}
Here ``links" is defined by the tuple $(\vec s, \vec e_i)$ and ``plaq" is defined by $(\vec s, \vec e_i, \vec e_j)$ where $i>j$.  We omit for convenience the extensive, constant contribution to the magnetic Hamiltonian which makes the full Hamiltonian positive-definite. In the strong coupling regime, $\hat H \to \hat H_0$, the eigenstates of which are the electric basis of the link states. The eigenvalues are proportional to the sum over the lattice of the $SU(3)$ quadratic Casimir:
\begin{align}
  \hat E^2 \ket{p,q} = \frac{p^2 + q^2 + pq + 3p + 3q}{3}\ket{p,q}.
\end{align}
The unperturbed ground state is the product state of all links in the trivial representation, denoted $\ket {0^{(0)}}$,  with energy $E_0 = \braket{0^{(0)}|\hat H_0}{0^{(0)}} = 0$. As discussed in \cite{KogutSusskind:1975:KS_hamiltonian}, excitations generated by $\hat V$ are gauge-invariant excitations of a plaquette with higher dimensional irreps. Larger excitations can be built by repeated application. For finite $g$, the true ground state and excitations are perturbed. Since the ground state is nondegenerate, however, it is simple to find corrections to the ground state as a power series in $g^{-1}$ \cite{SakuraiNapolitano:2020:QM}:
\begin{align}
\begin{split}
\label{eq:OG_PT}
    \ket{0} &= \ket{0^{(0)}} + \lambda \ket{0^{(1)}} + \lambda^2 \ket{0^{(2)}} + ...\\
    &= \ket{0^{(0)}} + \lambda\sum_{k\not=0}\ket{k_1} \frac{\braket{k_1}{\hat V|0^{(0)}}}{E_0^{(0)} - E_{k_1}^{(0)}} + \lambda^2\sum_{k\not=0}\frac{\ket{k_2} \braket{k_2}{\hat V|k_1} \braket{k_1}{\hat V|0^{(0)}}}{(E_0^{(0)}-E_{k_1}^{(0)})(E_0^{(0)}-E_{k_2}^{(0)})} + ...
\end{split}\end{align}
where $\ket{0^{(i)}}$ compose state corrections to different orders of perturbation and the states $k_1$ and $k_2$ are lattice states in the unperturbed Hamiltonian corresponding to first-order and second-order excitations respectively. The summation excludes excitations which are identical to the unexcited ground state.

\begin{figure}
    \centering
    \includegraphics[width=1\linewidth]{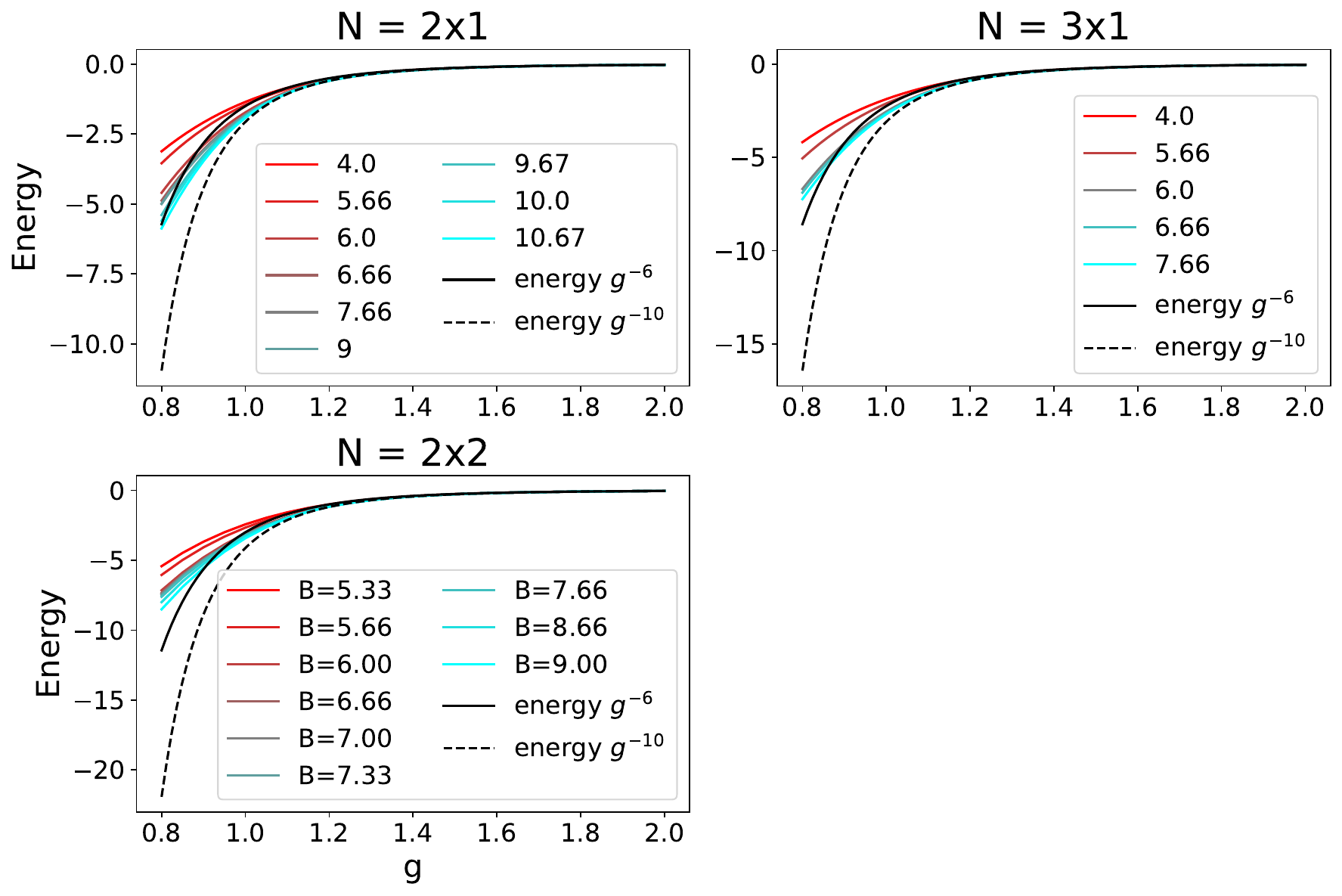}
    \caption{The ground state energy of various small lattices as a function of $g$. Colored curves correspond to exact diagonalization results with a range of $B$ truncations. Black solid and dashed curves correspond to strong-coupling lattice perturbation theory. Perturbation theory is seen to perform well down to $g\sim1$ and empirically convergence of the $B$ truncation is also observed. Note that as $B$ increases, the number of matrix elements for $\square$ and the number of $SU(3)$ irreps included increases as seen in Fig. \ref{fig:unsym_counts} and in Table \ref{tab:3_2d_B_singlets}. For example, in $d=3/2$, the {8} appears at $B=5.67$, the {6} appears at $B=6$ and the {15} appears at $B=9.67$.}
    \label{fig:PT_and_B}
\end{figure}

The first-order state correction can be simplified further since ${\braket{k_1|\hat V}{0^{(0)}} = 1}$ for all single plaquette excitations $k_1$, $E_0^{(0)} = 0$, and $E_k^{(0)} = \frac{8g^2}{3a^{d-2}}$. Using the shorthand $\ket{k_1} = \ket{p,s}$, where $p$ is the location of the plaquette and $s = \square,\square^\dagger$ defines the two ways it can be excited, we have the first order ground state correction:
\begin{align}\lambda \ket{0^{(1)}} &= \frac{3 a^{2d-6}}{8g^4} \sum_\text{plaq}\sum_{s=\square,\square^\dagger}\ket {p,s}.
\end{align}
In other words, the first-order correction is a uniform superposition of an excitation at each plaquette of the lattice. Since the state-correction is order $g^{-4}$, the true ground state converges very quickly to the unperturbed state for $g>1$ and small lattices. We may then want to consider just this first-order correction as $\ket 0 = \ket{0^{(0)}} + \lambda \ket{0^{(1)}}$. This state is not normalized, so one might include a factor of $Z^{-1} = \braket{0}{0}$ and define a normalized state 
\begin{align}
\ket 0_N = \left(1 + \frac{9N_Pa^{4d-12}}{32g^8}\right)^{-\frac12}\left(\ket{0^{(0)}} + \frac{3 a^{2d-6}}{8g^4} \sum_\text{plaq}\sum_{s=\square,\square^\dagger}\ket {p,s}\right)
\end{align}
where $N_P$ is the number of distinct plaquettes. Formally, however, this state is of mixed order in perturbation theory. The energy of this state is 
\begin{align}E_0(g) = -N_P\left(1 + \frac{9N_Pa^{4d-12}}{32g^8}\right)^{-1}\left(\frac{3a^{3d-10}}{4g^6} + \frac{9a^{5d-16}}{32g^{10}}\right).\label{eq:E0_eqn}
\end{align}
This computation is carried out in Appendix \ref{appendix:strong_PT}.
One might wonder why the lowest order correction to the ground state energy is $g^{-6}$. The reason is the energy shift is given by $\Delta_0 = \lambda \braket{0^{(0)}}{\hat V|0^{(0)}} + \lambda^2 \braket{0^{(0)}}{\hat V|0^{(1)}} + \mathcal O(\lambda^3)$, and since the magnetic perturbation $\hat V$ in the electric basis is zero on the diagonal, we see that the lowest order shift is proportional to $\lambda^2E_p^{-1} \sim g^{-6}$. Observe also that the energy of the normalized  corrected state is again not of fixed order in perturbation theory. Keeping only the lowest orders we have
\begin{align}
    E^{(1)}_0(g) = -N_P\left(\frac{3a^{3d-10}}{4g^6} + \frac{9a^{5d-16}}{32g^{10}}\right).
\end{align}
Normally the mixed-order $g^{-10}$ term in Eq.~(\ref{eq:E0_eqn}) for $E_0$ should be discarded, because the second-order state correction could introduce new corrections to order $g^{-10}$. However this does not happen. The $g^{-10}$ terms from $\ket{0^{(2)}}$ are $2\lambda^3\braket{0^{(2)}}{\hat H_0|0^{(1)}}$ and $2\lambda^3\braket{0^{(2)}}{\hat V|0^{(0)}}$, and $\hat H_0\ket{0^{(1)}} + \hat V\ket{0^{(0)}} = 0$, so these terms cancel. Thus the first-order state correction contains the two lowest order corrections entirely, and second-order state corrections would contribute at third lowest order, $g^{-14}$. This highlights the surprising effectiveness of the first-order state correction shown in Fig.~\ref{fig:PT_and_B}, where we compare the ground state energy in strong-coupling PT with exact diagonalization for a few small lattices. We also emphasize  this formula for the first-order state correction is independent of the lattice geometry and the truncation of irreps. Higher-order corrections will depend on these factors.

While we see good convergence of perturbation theory to the exact data even around $g=1.2$, for $g$ below unity the truncated series quickly becomes a poor approximation to the ground state. We also observe in Fig.~\ref{fig:PT_and_B} an empirical convergence of the ground state energy for fixed $g$ as the singlet energy cutoff $B$ is increased. The target for future quantum simulations is probably a lattice coupling around $g=1$, so we can hope that perturbation theory is a useful qualitative guide for more precise algorithms. Next we explore some ansatze that can provide a basis for VQE-type approaches to ground state preparation.

\subsection{Variational ans{\"a}tze}
\label{sec:em-emem-ansatz}

In the following it will be convenient to define the KS Hamiltonian as 
\begin{equation}
    \hat{H} = g^2 \hat{H}_E + \frac{1}{g^2} \hat{H}_B.
\end{equation}
and set $a=1$ for the remainder of the paper. We will examine different variational ansatz circuits  utilizing elements of the Trotterized time-evolution circuits developed in the \verb|ymcirc| package described below. These time-evolution circuits can be characterized as either evolving the electric part of the Hamiltonian or evolving the magnetic part of the Hamiltonian. We call the circuit for electric evolution circuit $E(t)$ and that for magnetic evolution $M(t)$, such that
\begin{subequations}
    \begin{align}
        &E(g^2t) = e^{-i g^2\hat{H}_Et} \\
        &M(t/g^2) = e^{-i \hat{H}_B t/g^2}.
    \end{align}
\end{subequations}

Our first ansatz circuit is build from successive applications of the ($E$)lectric and ($M$)agnetic circuits to the electric vacuum, replacing the time-evolution parameter $t$ with the generalized parameters $\{ \theta_i \}$ (we absorb any factors of $g$ into the parameters). The variational algorithm then attempts to find a local minimum for the expectation value of the energy in the state space spanned by the $\theta_i$. The successive applications of $E(\theta_i)$ and $M(\theta_j)$ in different combinations produces a family of operators, any fixed choice of which could be used as the ansatz. We focus first on the simplest combinations: (a) $E(\theta_2) M(\theta_1)$, and (b) $E(\theta_4) M(\theta_3) E(\theta_2) M(\theta_1)$. This has a flavor of the well-known Lanczos algorithm, where powers of the Hamiltonian are successively applied to an initial state \cite{Kirby2023exactefficient}. However here we allow the parameters $\theta_i$ to differ significantly from their values in the Hamiltonian and also do not include ``cross-terms" that would emerge like $EE$ or $MM$. The inspiration for these choices  lies in strong-coupling perturbation theory. 

Let us first consider first-order perturbation theory. As discussed in Section \ref{sec:PT} the (unnormalized) approximation to the ground state up to the first order in  $1/g$ is given by
\begin{equation} \label{eq:1pt-ground-state-prediction}
    \begin{split}
    \ket{0} &= \ket{0^{(0)}} - \frac{1}{g^2} \sum_{k_1} \frac{\bra{k_1}\hat{H}_B \ket{0^{(0)}}}{E_{k_1}} \ \ket{k_1} + \mathcal{O}(1/g^8) \\
    &= \ket{0^{(0)}} + \frac{3}{8g^4} \sum_{k_1} \ket{k_1} + \mathcal{O}(1/g^8)
    \end{split}
\end{equation}
where the $\ket{k_1}$ states refer to single-plaquette excitations of the lattice and $E_{k_1} = 8/3 g^2$  (note that this value is obtained by calculating the value of $H_E$ for all the excited links in the plaquette). Now, let us consider the parametric operator $M(\theta_1)$. In the strong-coupling limit when $\theta_1$ is sufficiently small, we can perturbatively expand the action of $M$ on the electric vacuum such that
\begin{equation}
    M(\theta_1) \ket{0^{(0)}} = \ket{0^{(0)}} - i \theta_1 \hat{H}_B \ket{0^{(0)}} + \mathcal{O} (1/g^8).
\end{equation}
Because the action of the plaquette operator (of which $\hat{H}_B$ is made up) leads to exactly the single-plaquette excitations $\ket{k_1}$, the above equation can also be written as
\begin{equation}\label{eq:em-strong-coupling-expansion}
    M(\theta_1) \ket{0^{(0)}} = \ket{0^{(0)}} +i \theta_1 \sum_{k_1} \ket{k_1} + \mathcal{O} (\theta_1^2)
\end{equation}
Therefore, the form of the strong-coupling expansion for $M(\theta_1)$ matches with perturbation theory up to the first order. However, the coefficient of the first order term in the expansion of $M(\theta_1)$ has an extra $i$,  which cannot be reconciled with first-order perturbation theory unless $\theta_1$ is also imaginary. This is not allowed since $M(\theta_1)$ is unitary. To remedy the situation we introduce the operator $E(\theta_2)$ in Eq.~(\ref{eq:em-strong-coupling-expansion}) such that
\begin{equation}
    E(\theta_2) M(\theta_1) \ket{0^{(0)}} = \ket{0^{(0)}} + i \theta_1 e^{-8i \theta_2/3} \  \sum_{k_1} \ket{k_1} + \mathcal{O}(\theta_1^2).
\end{equation}
Comparing Eq.(~\ref{eq:1pt-ground-state-prediction}) and Eq.~(\ref{eq:em-strong-coupling-expansion}), we find that in order to match the first order perturbative expansion,
\begin{align}
    \theta_1 = 3/(8g^4),~~~~\theta_2 = 3 \pi/16.
\end{align} 
Notice that while $\theta_1$ is perturbative in the strong-coupling limit, the value of $\theta_2$ is not perturbative as $E(\theta_2)$ is introduced just to fix a relative phase.

\begin{figure}[!ht]
    \centering
    \includegraphics[width=0.8\linewidth]{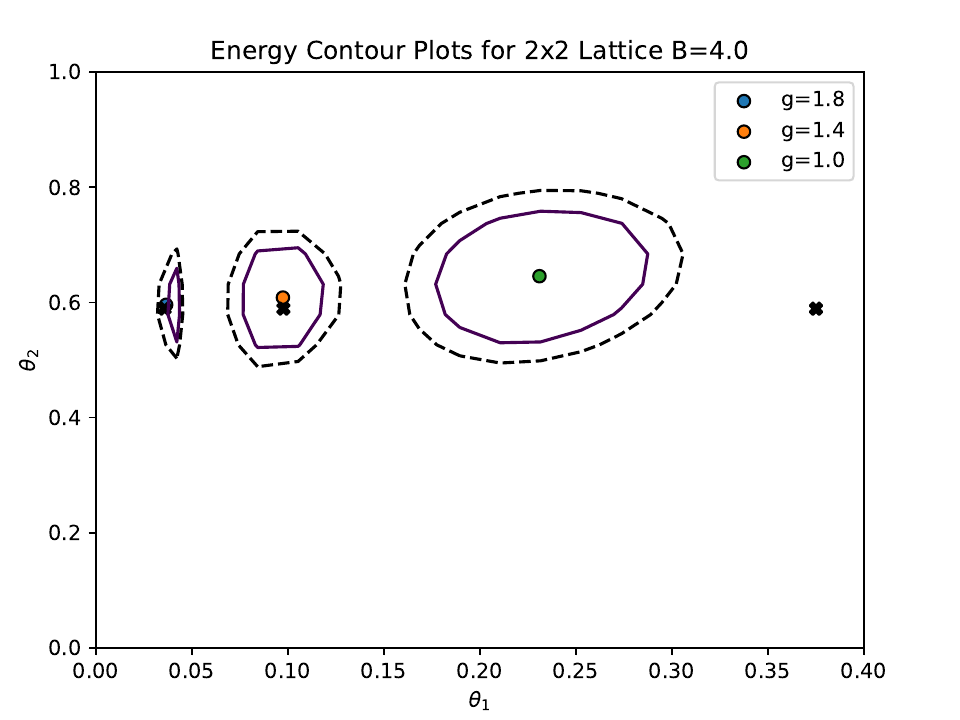}
    \caption{Energy contours for the EM ansatz states on the $2\times 2$ plaquette lattice with periodic boundary conditions and $B=4.0$ energy truncation. The black crosses denote the prediction of strong-coupling PT, while the colored dots denote the local minima. Solid and dashed contours denote 6\% and 10\% deviations from the minima, respectively.}
    \label{fig:contour-plot}
\end{figure}

Therefore, the parametric state $E(\theta_2) M(\theta_1) \ket{0^{(0)}}$ is expected to be a good ansatz in the strong-coupling region. For brevity, we will call this  the EM ansatz. In Fig.~\ref{fig:contour-plot}, we plot the 90\% and 94\% energy contours in the $\{\theta_1,\theta_2\}$ parameter space, along with the parameters obtained in the
minimization algorithm, for the $d=2$ lattice with $2\times 2$ plaquettes with couplings $g=1.0$, $g=1.4$, and $g=1.8$ in the $B=4.0$ energy truncation.  For $g=1.8$ and $g=1.4$, the agreement of the parameters obtained by VQE and that predicted by strong-coupling perturbation theory is excellent, whereas that for $g=1.0$ the fit deviates from PT.

To improve on the EM ansatz, we turn to second-order perturbation theory. The action of $M(\theta_1)$ on the electric vacuum up to second order in the parameter is given as 
\begin{equation}
    M(\theta_1) \ket{0^{(0)}} = \ket{0^{(0)}} + i \theta_1 \sum_{k_1} \ket{k_1} - \frac{\theta_1^2}{2} \sum_{\mathcal{E}=0}^{20/3} \sum_{\{E(k_2) = \mathcal{E}\}} \sigma(k_2) \ket{k_2,\mathcal E} + \mathcal{O}(\theta_1^3).
\end{equation}
The coefficient $\sigma(k_2)$ is the matrix element denoting the amplitude of this transition, i.e., $
    \sigma(k_2) = \bra{k_2} \hat{H}_B^2 \ket{0^{(0)}}
$. The $\ket{k_2, \mathcal E}$ are the two-plaquette excitations obtained by successively applying the plaquette operators, i.e. $\ket{k_2,\mathcal E} = - \hat{H}_B \ket{k_1}$,
and with an index given by their $H_E$ eigenvalue $\mathcal E$.
Accordingly, the sum over all the two-plaquette excitations is split over all the distinct $\mathcal{E}$ that result from a two-plaquette excitation (see Table \ref{tab:Givens_operators}), and  over all the two-plaquette excitations that have the same $\mathcal{E}$. Applying the $E(\theta_2)$ operator on both sides, we get
\begin{equation}
    E(\theta_2)M(\theta_1) \ket{0^{(0)}} = \ket{0^{(0)}} + i \theta_1 e^{-8i\theta_2/3} \sum_{k_1} \ket{k_1} - \frac{\theta_1^2}{2} \sum_{\mathcal{E}=0}^{20/3} e^{- i \mathcal{E} \theta_2} \sum_{\{E(k_2) = \mathcal{E}\}} \sigma(k_2) \ket{k_2,\mathcal E}
\end{equation}
Applying the EM combination one more time, we get
\begin{equation}\label{eq:emem-strong-coupling-expansion}
\begin{aligned} 
    E(\theta_4)M(\theta_3) E(\theta_2) M(\theta_1) \ket{0^{(0)}} = \ket{0^{(0)}} + i e^{-8i\theta_4/3} (\theta_1 e^{-8i\theta_2/3} + \theta_3) \sum_{k_1} \ket{k_1} & \\ -\sum_{\mathcal{E}=0}^{20/3} e^{-i\mathcal{E}\theta_4}\left(\frac{\theta_1^2 e^{- i \mathcal{E} \theta_2}}{2} + \frac{\theta_3^2}{2} + \theta_1 \theta_3 e^{-8i \theta_2/3} \right) \sum_{\{E(k_2) = \mathcal{E}\}} \sigma(k_2) \ket{k_2,\mathcal E}.
\end{aligned}
\end{equation}
In strong-coupling PT, Eq.~(\ref{eq:OG_PT}) tells us
\begin{equation}
    \ket{0^{(2)}} = \frac{3}{8g^8} \sum_{\mathcal{E}=8/3}^{20/3} \frac{1}{\mathcal{E}} \sum_{\{E(k_2) = \mathcal{E}\}} \sigma(k_2) \ket{k_2,\mathcal E}
\end{equation}
so, up to the second order, the perturbative ground state is given by
\begin{equation}\label{eq:2pt-ground-state-prediction}
    \ket{0} = \left(1 - \frac{9 N_P}{64 g^8} \right) \ket{0^{(0)}} + \frac{3}{8g^4} \sum_{k_1} \ket{k_1} + \frac{3}{8g^8} \sum_{\mathcal{E}=8/3}^{20/3} \frac{1}{\mathcal{E}} \sum_{\{E(k_2) = \mathcal{E}\}} \sigma(k_2) \ket{k_2,\mathcal E}
\end{equation}
where $N_P$ is the number of plaquettes in the lattice. Thus, while the expression for the perturbative expansion of the EMEM ansatz matches qualitatively with the strong-coupling expansion of the ground state, the former does not have enough free parameters to match the latter exactly: matching Eqs.~(\ref{eq:emem-strong-coupling-expansion}) and (\ref{eq:2pt-ground-state-prediction}) amounts to nine equations, corresponding to the number of sets of single and double plaquette excitations that have distinct electric energies $\mathcal{E}$, whereas we only have four independent parameters in the EMEM ansatz.  In smaller $B$ truncations the higher $\mathcal{E}$ excitations may be prohibited, so EMEM is a more effective ansatz for such truncations.

The gate depth cost of the EM and EMEM ans{\"a}tze scale essentially as one and two Trotter steps, respectively. (They can also be thought of as analogous to QAOA ans{\"a}tze.) We can do better, and improve on the ansatz quality, by considering circuits with more variables. 

In our approach to time evolution each distinct magnetic transition is implemented sequentially, on each plaquette, by two-level unitaries known as Givens rotations. 
Now let us consider an ansatz again based on the time evolution circuits, but for which each magnetic Givens rotation is assigned an independent parameter. Below we refer to ans{\"a}tze of this type -- and restrictions to subsets of the most important two-level transitions -- as ``multi-Givens".

We create an ansatz with parameters over 8 different classes of Givens rotations such as given in Table \ref{tab:Givens_operators}. The operator $\hat{G}_1$ jumps the electric vacuum into a single-plaquette excitation. It is like the magnetic Hamiltonian, restricted to just those matrix elements that are connected to the electric vacuum. Its effect is given  (up to second order in the parameter $\theta_1$) by 
\begin{equation}\label{eq:givens_g1_on_vacuum}
    e^{-i \hat{G}_1 \theta_1} \ket{0^{(0)}} = \left(1- \theta_1^2 N_P \right)\ket{0^{(0)}} + i \theta_1 \sum_{k_1} \ket{k_1} - \frac{\theta_1^2}{2} \sum_{k_1,k_1'} \ket{k_1}\otimes \ket{k_1'}
\end{equation}
where $k_1$ and $k_1'$ are single-plaquette excitations on disjoint plaquettes, i.e. plaquettes that share neither a corner nor edge. The rest of the Givens operators connect non-vacuum plaquette states, and so they only act nontrivially once the plaquette state (or more precisely, the state of the plaquette and its control links) is excited. Thus $\hat G_j \ket{0^{(0)}} = 0$ for $j \not=1$. Following $\hat G_1$ their combined effect on the electric vacuum is given (again up to second order in the parameters) by:
\begin{multline}
    \prod_{j=2}^8 e^{-i \sigma(k_j)\hat{G}_j \theta_j} \, e^{-i \hat{G}_1 \theta_1}\ket{0^{(0)}} = \left(1- \theta_1^2 N_P \right)\ket{0^{(0)}} + i \theta_1 \sum_{k_1} \ket{k_1} \\ -\sum_{j=2}^{8} \theta_1 \theta_j \sum_{\{E(k_2) = \mathcal{E} (j)\}} \sigma(k_2) \ket{k_2,\mathcal E} - \frac{\theta_1^2}{2} \sum_{k_1,k_1'} \ket{k_1}\otimes \ket{k_1'}.
\end{multline}
In order to fix the relative phases between the zeroth, first and second orders, we introduce an electric operator that adds a phase to the single plaquette excitation. Such an operator can be built by a multi-control phase gate wherein the controls are given by the active links of a single plaquette excitation. We will call this phase operator $\hat{P}_1$, with associated parameter $\tilde{\theta}_1$. After applying the exponent of this operator to both sides, we obtain
\begin{multline}
    e^{-8i \hat{P}_1 \tilde{\theta}_1/3} \prod_{i=1}^8 e^{-i \sigma(k_i) \hat{G}_i \theta_i} \ket{0^{(0)}} = (1-\theta_1^2 N_P) \ket{0^{(0)}} + i \theta_1 e^{-8i \tilde{\theta}_1/3} \sum_{k_1} \ket{k_1} \\- \frac{\theta_1^2}{2} e^{-16i\tilde{\theta}_1/3} \sum_{k_1,k_1'} \ket{k_1} \otimes \ket{k_1'} - \sum_{j=2}^{8} \theta_1 \theta_j e^{-16i \tilde{\theta}_1 \delta_{j4}/3} \sum_{\{E(k_2) = \mathcal{E}(j)\}} \sigma(k_2) \ket{k_2,\mathcal E} .
\end{multline}
Comparing this to second order state in Eq.~(\ref{eq:2pt-ground-state-prediction}), we find that strong-coupling PT predicts
\begin{equation}
\begin{aligned}
    \theta_1 &= \frac{3}{8 g^4} + \frac{9}{64 g^8} \; ,& \theta_2 &= - \frac{1}{4 g^4} \; , & \theta_3 &= - \frac{3}{14 g^4} \; , \\
     \theta_4 &= \frac{3}{16 g^4} \; , & \theta_5 &= -\frac{6}{33 g^4} \; , & \theta_6 &= - \frac{3}{17 g^4} \; , \\
     \theta_7 &= - \frac{1}{6 g^4} \; , & \theta_8 &= - \frac{3}{20 g^4} \; , & \tilde{\theta}_1 &= \frac{3\pi}{16} \; . \\
\end{aligned}
\end{equation}
Notice that even though there are three equations for $\theta_1$ when we match the zeroth, first orders, and the disjoint plaquette parts of the ansatz and perturbation theory, the solution $\theta_1 = 3/(8g^4) + 9/(64g^8)$ is consistent with all three orders.\footnote{Observe in Eqn. \ref{eq:2pt-ground-state-prediction} that there are double-excitations that look like single-excitations. This follows from the fact that $1 = \braket{0^{(0)}}{\square^3 |0^{(0)}} = \braket{\square^\dagger 0^{(0)}}{\square^2 0^{(0)}}$. $\theta_1$ includes these double-excitation terms and thus has this higher-order contribution.}

\begin{table}[!ht]
\centering
\begin{tabular}{ l  >{\centering\arraybackslash}m{6cm} l}
\toprule
Plaquette excitation & $\mathcal{E}$ & Givens Operators \\
\midrule
&\centering Single Plaquette Excitations & \\
\midrule
\begin{tikzpicture}[baseline={(current bounding box.center)}, scale=0.5, every node/.style={scale=0.5}]
  \draw (0,0)  -| (2,2) 
    node[pos=0.25,below] {$3$} 
    node[pos=0.75,right] {$3$}
    -| (0,0)
    node[pos=0.25,above] {$\bar{3}$}
    node[pos=0.75,left] {$\bar{3}$};
  \end{tikzpicture} $+$ h.c. & $8/3$ & $\hat{G}_{1}$, $\hat{P}_1$ \\
  \toprule
  &\centering Two Plaquette Excitations& \\ 
  \midrule
\begin{tikzpicture}[baseline={(current bounding box.center)}, scale=0.5, every node/.style={scale=0.5}]
  \draw (0,0)  -| (2,2) 
    node[pos=0.25,below] {$3$} 
    node[pos=0.75,right] {$1$}
    -| (0,0)
    node[pos=0.25,above] {$\bar{3}$}
    node[pos=0.75,left] {$\bar{3}$};
  \draw (2,0)  -| (4,2) 
    node[pos=0.25,below] {$3$} 
    node[pos=0.75,right] {$3$}
    -| (2,0)
    node[pos=0.25,above] {$\bar{3}$};
  \end{tikzpicture} $+$ h.c. & $12/3$ & $\hat{G}_2$ \\
  \begin{tikzpicture}[baseline={(current bounding box.center)}, scale=0.5, every node/.style={scale=0.5}]
  \draw (0,0)  -| (2,2) 
    node[pos=0.25,below] {$3$} 
    node[pos=0.75,right] {$\bar{3}$}
    -| (0,0)
    node[pos=0.25,above] {$\bar{3}$}
    node[pos=0.75,left] {$\bar{3}$};
  \draw (2,0)  -| (4,2) 
    node[pos=0.25,below] {$\bar{3}$} 
    node[pos=0.75,right] {$\bar{3}$}
    -| (2,0)
    node[pos=0.25,above] {$3$};
  \end{tikzpicture} $+$ h.c. & $14/3$ & $\hat{G}_3$ \\
\begin{tikzpicture}[baseline={(current bounding box.center)}, scale=0.5, every node/.style={scale=0.5}]
  \draw (0,0)  -| (2,2) 
    node[pos=0.25,below] {$3$} 
    node[pos=0.75,right] {$3$}
    -| (0,0)
    node[pos=0.25,above] {$\bar{3}$}
    node[pos=0.75,left] {$\bar{3}$};
  \draw (2.2,2.2)  -| (4,4) 
    node[pos=0.25,below] {$3$} 
    node[pos=0.75,right] {$3$}
    -| (2.2,2.2)
    node[pos=0.25,above] {$\bar{3}$}
    node[pos=0.75,left] {$\bar{3}$};
\end{tikzpicture} $+$ perms. & $16/3$ & $\hat{G}_4$ \\
\begin{tikzpicture}[baseline={(current bounding box.center)}, scale=0.5, every node/.style={scale=0.5}]
  \draw (0,0)  -| (2,2) 
    node[pos=0.25,below] {$3$} 
    node[pos=0.75,right] {$8$}
    -| (0,0)
    node[pos=0.25,above] {$\bar{3}$}
    node[pos=0.75,left] {$\bar{3}$};
  \draw (2,0)  -| (4,2) 
    node[pos=0.25,below] {$3$} 
    node[pos=0.75,right] {$3$}
    -| (2,0)
    node[pos=0.25,above] {$\bar{3}$};
  \end{tikzpicture} $+$ h.c. & $11/2$ & $\hat{G}_5$ \\
\begin{tikzpicture}[baseline={(current bounding box.center)}, scale=0.5, every node/.style={scale=0.5}]
  \draw (0,0)  -| (2,2) 
    node[pos=0.25,below] {$3$} 
    node[pos=0.75,right] {$6$}
    -| (0,0)
    node[pos=0.25,above] {$\bar{3}$}
    node[pos=0.75,left] {$\bar{3}$};
  \draw (2,0)  -| (4,2) 
    node[pos=0.25,below] {$\bar{3}$} 
    node[pos=0.75,right] {$\bar{3}$}
    -| (2,0)
    node[pos=0.25,above] {$3$};
  \end{tikzpicture} $+$ h.c. & $17/3$ & $\hat{G}_6$ \\
\begin{tikzpicture}[baseline={(current bounding box.center)}, scale=0.5, every node/.style={scale=0.5}]
  \draw (0,0)  -| (2,2) 
    node[pos=0.25,below] {$8$} 
    node[pos=0.75,right] {$8$}
    -| (0,0)
    node[pos=0.25,above] {$8$}
    node[pos=0.75,left] {$8$};
  \end{tikzpicture} $+$ h.c. & $18/3$ & $\hat{G}_{7}$ \\
  \begin{tikzpicture}[baseline={(current bounding box.center)}, scale=0.5, every node/.style={scale=0.5}]
  \draw (0,0)  -| (2,2) 
    node[pos=0.25,below] {$6$} 
    node[pos=0.75,right] {$6$}
    -| (0,0)
    node[pos=0.25,above] {$\bar{6}$}
    node[pos=0.75,left] {$\bar{6}$};
  \end{tikzpicture} $+$ h.c. & $20/3$ & $\hat{G}_{8}$ \\
\bottomrule
\end{tabular}
\caption{Givens operator groups. Each group $\hat{G}_i$ denotes Givens rotations that have the same electric energy. The notation is such that $\hat G_{j}$ for $j>1$ refers to single-plaquette magnetic transitions that connect non-electric-vacuum states, one of which is a single-plaquette excitation. Thus the h.c. in the loop of {8} is a reminder that it can be formed either from an application of $\square$ then $\square^\dagger$, or $\square^\dagger$ then $\square$.\label{tab:Givens_operators}}
\end{table}

The above analysis holds for arbitrary large-scale lattices in general dimensions. Moreover, selectively choosing Givens operators based on matching with perturbation theory  can  give a systematic way of improving the accuracy of the  ansatz up to a given order in $1/g$.  However, the perturbative calculations for multi-Givens might not always hold for small periodic lattices because control links can wrap around in such lattices and act as controls for multiple vertices. 

We also demonstrate in later sections that selecting Givens rotations based on how ``important" they are in creating the ground state can lead to significant savings in the gate depth of the state preparation circuits.

\subsection{Adiabatic and Hybrid VQE-Adiabatic preparation}
Another approach to state preparation is the adiabatic method. In the present context, this entails preparing the strong coupling electric vacuum (trivially, since it is the all-zeros computational basis state), then time-evolving over many small Trotter steps while the coupling is gradually lowered to a target value. 

We will see that for small lattices, both variational ans{\"a}tze and adiabatic methods generate accurate approximations to the ground state. The variational method has an up-front cost driven by the iterative procedure to find the parameters for the given ansatz with maximal ground state overlap. Once these parameters have been determined, the state preparation then requires only (relatively) shallow circuits, of a depth comparable to one (for EM) or two (for EMEM) steps of Trotter evolution, and significantly less for the multi-Givens ansatz. At smaller couplings  we observe that the ability of the ansatz states to capture the true ground state degrades somewhat.

On the other hand, the adiabatic algorithm maintains a strong overlap with the true ground state roughly independent of the value of $g$ (see e.g.\ Fig.~\ref{fig:1cubeFidelities} below). However, since the evolution begins at large $g$, the preparation circuit becomes  deep as $g$ is decreased. This makes state preparation inefficient, especially since the evaluation of ground state observables will generically require preparing the ground state repeatedly to build up measurement statistics.

To minimize state depth while maintaining strong ground state overlap irrespective of $g$, we consider a hybrid strategy that uses the optimized VQE circuit for $g$ up to some threshold value and then ``switches on'' adiabatic preparation when simulating at $g$ below that threshold.\footnote{A similar strategy has previously been used in initial state preparation of scalar lattice $\phi^4$ theory \cite{li-2023-simulatin}.} In cases where the true ground state is known, one can compute $1-F$ as a function of $g$ for both methods. The adiabatic runs can now be considered part of the ``up front'' cost, similar to the  optimization runs. The threshold can be taken to be a $g$ near where the adiabatic and variational ansatz curves cross (see e.g.\ Figs.~\ref{fig:1cubeFidelities} and \ref{fig:cubic_fidelities} below).

In the more interesting case when the true ground state is not known, a threshold value must be inferred in order to use the hybrid approach. One possibility is to choose a threshold motivated by strong-coupling PT and then empirically measure the sensitivity of the adiabatically-prepared state at lower $g$ to small variations in the threshold.

\section{Ground state preparation: Results}
It is straightforward to test state preparation algorithms in classical simulations of small lattice systems. In the following subsections, we compare variational and adiabatic methods for ground state preparation of various test lattices in $d=3/2$, $d=2$, and $d=3$, and comment on which preparation methods are most effective in each of these test cases. We assess the relative efficacy of these methods by comparing the ground state energy obtained with exact diagonalization for various $B$ truncations to the expectation value of the energy in the best-fit ansatz states. We also compute the fidelity of the ansatz states (defined by comparison to exact diagonalization) and contrast these results with those obtained by adiabatic state preparation.

\subsection{2-plaquette Chain Lattice}\label{sec:2-plaq-results}
 We begin with the 2-plaquette chain, varying the parameters of the EM and EMEM ans{\"a}tze defined in Sec.\,(\ref{sec:em-emem-ansatz}) until a (local) minima for the expectation value of the Kogut-Susskind Hamiltonian is obtained. The Broyden-Fletcher-Goldfarb-Shanno algorithm that is built into SciPy's \texttt{minimize} method was used to reach this minima up to a specified precision. The ability of any minimization algorithm to find the global minimum depends on the values of the initialization parameters fed into the algorithm. In our particular case, it is easy to guess the values of initialization parameters that should be fed into the algorithm for larger values of the coupling constants, where we can use perturbation theory directly (see Sec.\,(\ref{sec:em-emem-ansatz})). However, perturbation theory predictions fail to provide good initialization parameters at weak coupling. Therefore we use a  modified variational algorithm in which estimates for initialization parameters  are obtained by successively running the optimization starting from a fixed strong coupling point and adiabatically decreasing the value of the coupling constant, recursively feeding the optimal parameters obtained for $g_n$ as the initialization parameters for $g_{n+1}$ until the target coupling is reached.

\begin{figure}[!ht]
    \centering
     \begin{subfigure}{0.48\textwidth}
         \includegraphics[width=\textwidth]{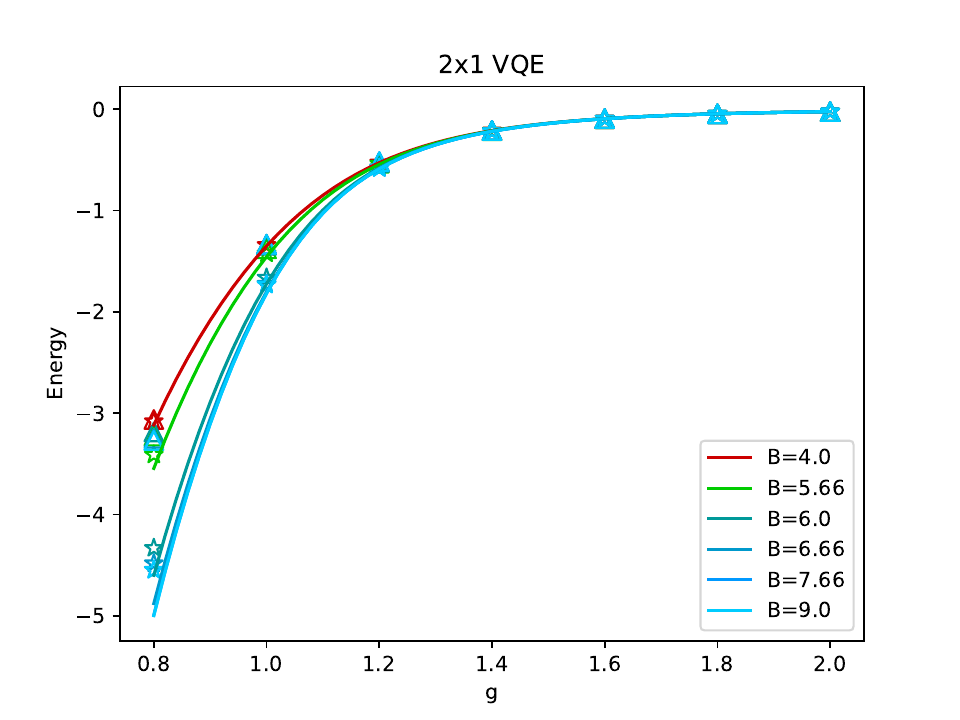}
         \caption{}
         \label{fig:2x1_vqe_energies}
     \end{subfigure}
     \begin{subfigure}{0.48\textwidth}
         \includegraphics[width=\textwidth]{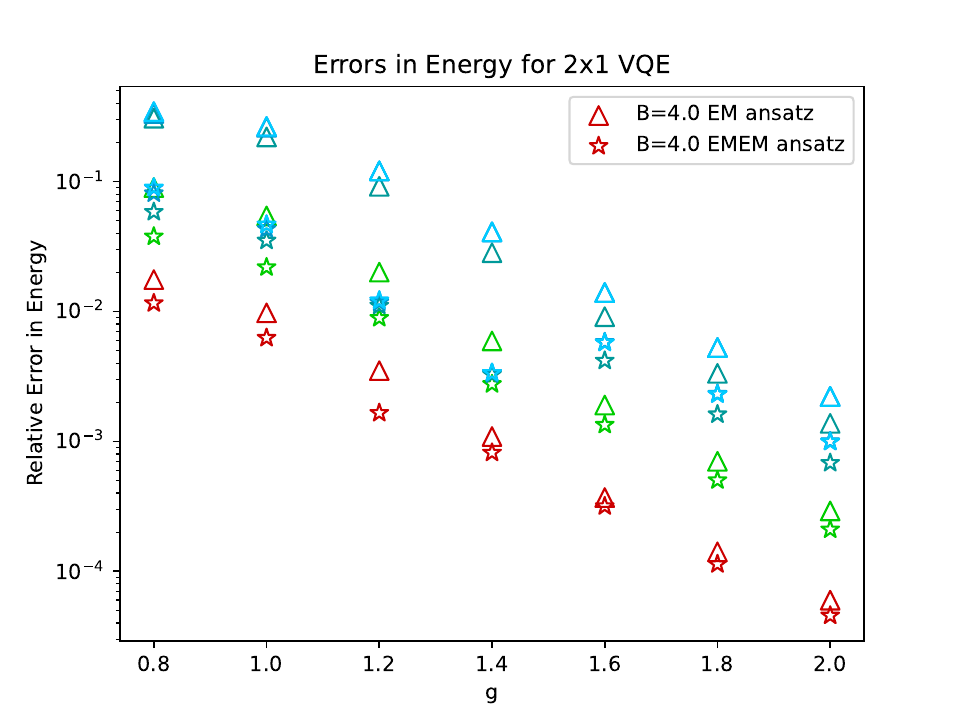}
         \caption{}
         \label{fig:2x1_vqe_relative_error}
     \end{subfigure}
     \caption{(a) Comparison of ground state energies obtained by VQE using the EM and EMEM ans{\"a}tze to exact diagonalization for different values of energy truncation $B$. The solid lines depict the exact diagonalization energies, the triangles depict the energies obtained using the EM ansatz, and the stars depict the energies obtained using the EMEM ansatz. (b) Relative error in energies obtained by VQE with respect to exact diagonalization. The markers for EM and EMEM are the same as (a).}
     \label{fig:2x1_VQE}
\end{figure}

We compare the ground state energies obtained by the variational algorithm with those obtained by exact diagonalization in Fig.~\ref{fig:2x1_VQE}. We see that as the site singlet energy cutoff $B$ is increased, the relative error between the energies of VQE with EM/EMEM ansatz and exact diagonalization energies increases, indicating that a better ansatz with more parameters is needed for higher B-cutoffs. However, EMEM improves significantly on EM, and remains good to better than 10\% even down to $g=0.8$ for $B=9$ (which includes, for example, some singlets involving $6,\bar6,8$.)

From Fig.~\ref{fig:2x1_VQE}, it is apparent that the relatively simple EM and EMEM ans\"{a}tze become insufficient in describing the properties of the ground state for higher $B$-truncations. This is because higher $B$-truncations allow for more energetic  states to mix into the ground state, the amplitudes for which need to be fitted with more parameters than are included in the EM or EMEM ans\"{a}tze. Therefore, more sophisticated ans\"{a}tze with more parameters like the multi-Givens ansatz become more useful in such cases. However, we defer the implementation of this ansatz to bigger lattices, such as the cubic lattice with open boundary conditions, since perturbative calculations of multi-Givens parameters such as the one presented in Sec.~\ref{sec:em-emem-ansatz} are plagued with special-case annoyances produced by repeated control links in small lattices with periodic boundary conditions, which do not generalize to arbitrarily large lattices.

\subsection{Cubic Lattice}\label{sec:cube-results}
The modified variational optimization procedure described in Sec.\,(\ref{sec:2-plaq-results}) can be implemented on larger lattices. The three dimensional cube with open boundary conditions is relatively easy to study in the $T_1$ truncation (equivalent to $B=4.0$ for this lattice) because the circuits are shallow and the qubit count is limited. In addition to the EM and EMEM ans{\"a}tze used in Sec.\,(\ref{sec:2-plaq-results}), we test the ``multi-Givens" ansatz described earlier. In this ansatz a restricted collection of magnetic transitions is utilized, each with a different parameter. Here our choice of this collection is motivated by strong coupling PT; in practice we perform a fit at $g=2$ with all rotations given a free parameter, then select only those rotations for which the fit assigns a nonzero parameter. Without this cut, the CX gate count is $\sim 26{,}000$, but removing the relatively less important Givens rotations reduces this count to  $\sim8{,}000$. 

In larger lattices, finding the minima of the energy function by iterating over the ansatz circuits might become a computationally expensive task. One approach is to weave together circuits for small lattices into larger ones by successive applications of VQE~\cite{Ciavarella:2022:su3_state_prep}. Testing the weaving approach on small enough systems to simulate classically would be an interesting direction for future study.

\begin{figure}[!ht]
    \centering
    \includegraphics[width=0.6\linewidth]{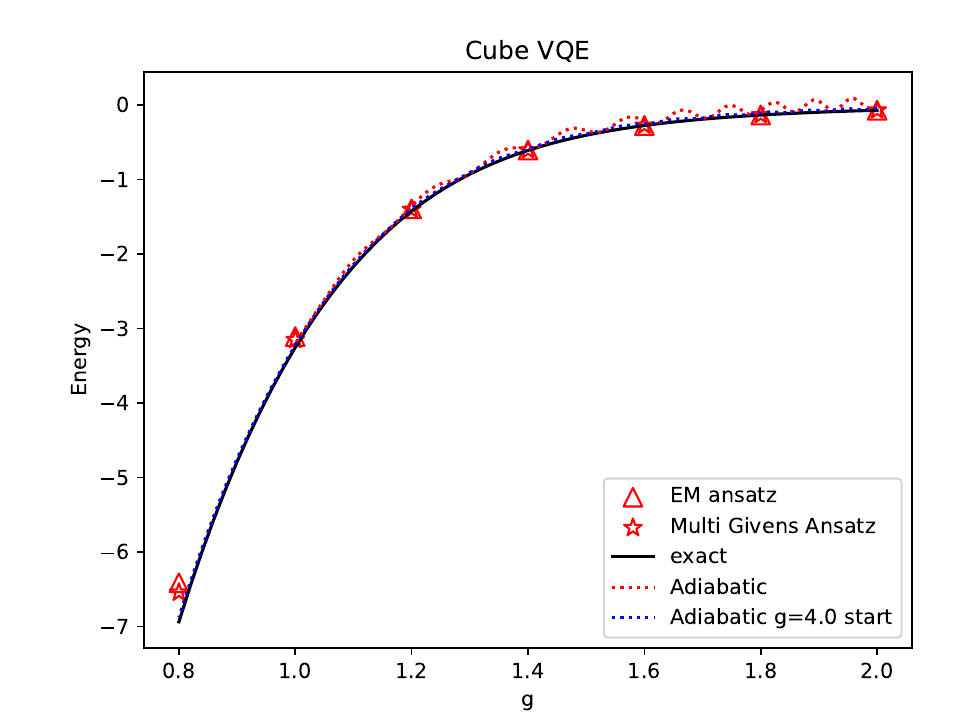}
    \caption{Comparison of cube lattice ground-state energy obtained via variational methods with the EM and multi-Givens ans{\"a}tze described in the text; exact diagonalization; and adiabatic state preparation starting from $g=4$ (blue) and $g=2$ (red) respectively, with adiabatic step sizes of $\Delta g = 0.01$ in the $B=4.0$ (or equivalently, the $T_1$) truncation.}
    \label{fig:cubic_adiabatic}
\end{figure}

 In Fig.~\ref{fig:cubic_adiabatic} we compare the energies and the relative error in energy obtained for the cube lattice by different methods of state preparation. We see that all methods work fairly well down to $g\sim 1$. At lower $g$, adiabatic state preparation outperforms our variational ans{\"a}tze. At larger $g$, the quality of the adiabatic method depends on the starting point. To illustrate this, we include
 two adiabatic curves, corresponding to different initial values of $g$. Starting the adiabatic evolution from the electric vacuum in the far strong coupling regime leads to a more stable evolution than starting at $g=2.0$, since the electric vacuum is a better approximation to the ground state in the strong coupling regime. However, starting the adiabatic evolution from a higher value in $g$ has a tradeoff in gate counts. In the figure,  $\sim 300$ Trotter steps were used to adiabatically evolve the electric vacuum to $g=1.0$ from $g=4.0$, and $\sim 100$ Trotter steps were used to evolve to $g=1.0$ from $g=2.0$. Of course, this can be improved by using a better perturbative starting point, or by a hybrid adiabatic-variational method discussed in the next subsection.

\begin{figure}[!ht]
    \centering
    \includegraphics[width=0.6\linewidth]{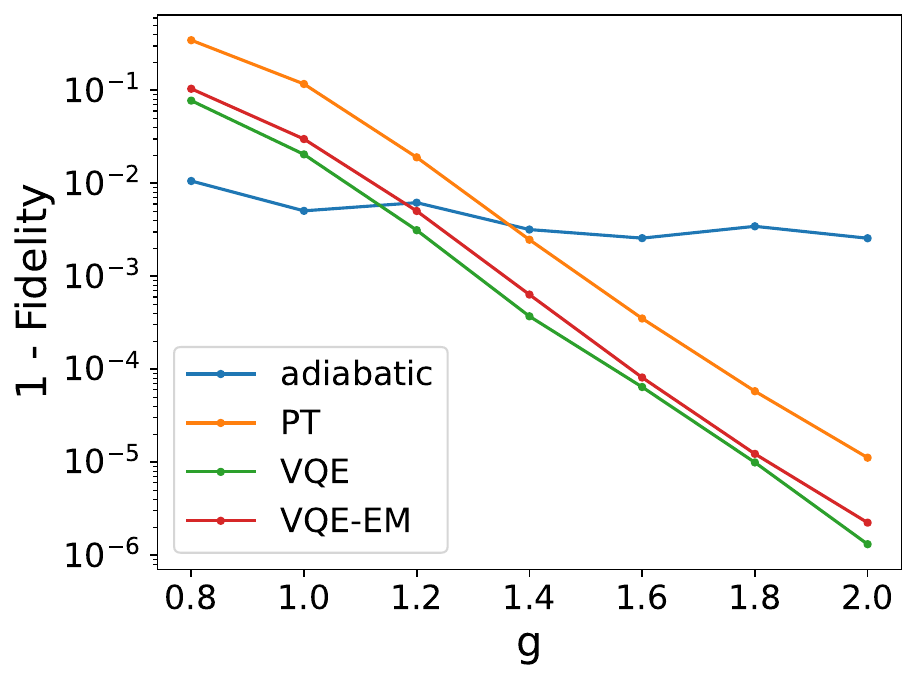}
    \caption{Fidelity of the approximate ground state prepared using the adiabatic method, the variational method with EM ansatz (``VQE-EM"), the variational method with the multi-Givens ansatz (``VQE"), and lowest order perturbation theory for the cubic lattice. Here fidelity refers to $|\braket{\psi_\text{ED}}{\psi}|^2$ where $\psi_\text{ED}$ is the ground state obtained via exact diagonalization. The most interesting result is that the relatively shallow multi-Givens ansatz achieves percent-level fidelity at $g\sim 1$.}
    \label{fig:1cubeFidelities}
\end{figure}

In Fig.~\ref{fig:1cubeFidelities} we show the state fidelity, the probability of obtaining the true ground state (defined as statevector obtained by exact diagonalization) in the prepared approximate ground state. Small-step adiabatic preparation achieves the best results at small $g$, and is systematically improvable by either increasing the starting value of $g$ or improving the starting state. However, the multi-Givens ansatz, based on the most important magnetic transitions at strong coupling, achieves percent-level fidelity down to $g\sim 1.0$.

\begin{figure}[!ht]
\centering
\includegraphics[width=0.6\textwidth]{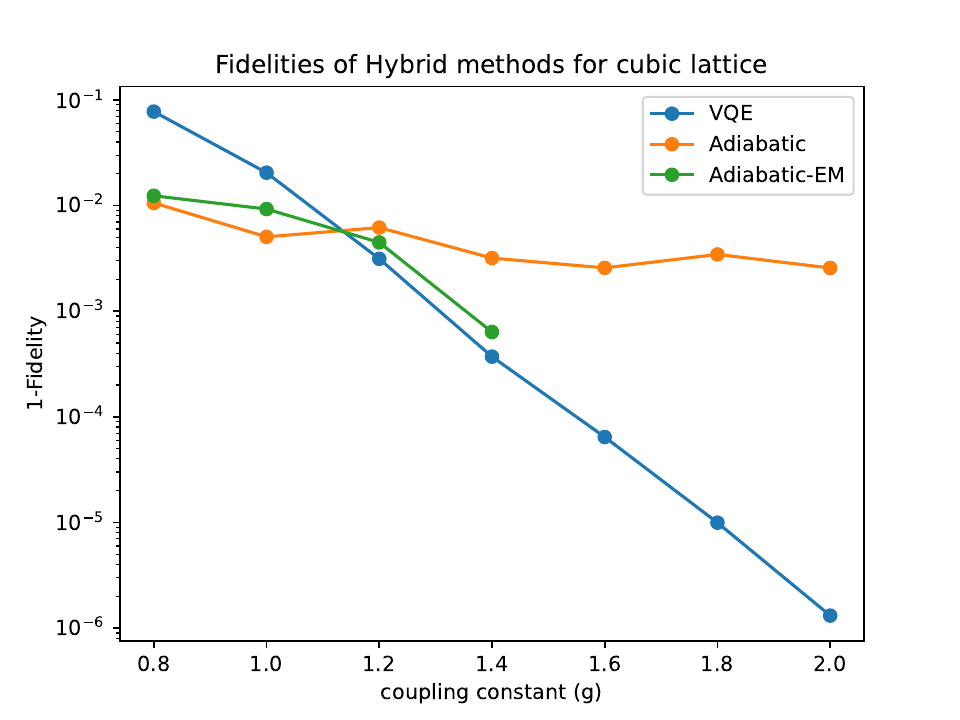}
\caption{Fidelities for different state preparation methods on the cube. The hybrid method Adiabatic-EM first uses the optimized VQE-EM circuit from $g=1.4$ and then use the adiabatic algorithm to compute the ground state at smaller $g$.  
\label{fig:cubic_fidelities}
}
\end{figure}

Now let us demonstrate the hybrid adiabatic-VQE method on the cube lattice. Fig.~\ref{fig:cubic_fidelities} repeats the fidelities, this time comparing the fidelities obtained by adopting a hybrid ``adiabatic-EM" approach to that obtained by the multi-Givens and adiabatic approach described in the previous section. The hybrid adiabatic state preparation is turned on around $g=1.4$, for which the VQE preparation circuits have a $1-F \sim 10^{-3}$. The $1-F$ value of the hybrid adiabatic evolution is very similar to that of adiabatic evolution at $g=0.8$, but the circuit depth for the hybrid approach is significantly lower, as the hybrid approach only needs $\sim 25$ Trotter steps to reach $g=0.8$, whereas the adiabatic approach needs $\sim 300$ steps. Therefore, the hybrid approach affords fidelities comparable to the traditional adiabatic approach at a much lower gate  cost. 


\subsection{\texorpdfstring{$2\times 2$ plaquettes}{2x2 plaquettes}}

\begin{figure}[!ht]
    \centering
    \begin{subfigure}{0.48\textwidth}
        \includegraphics[width=\linewidth]{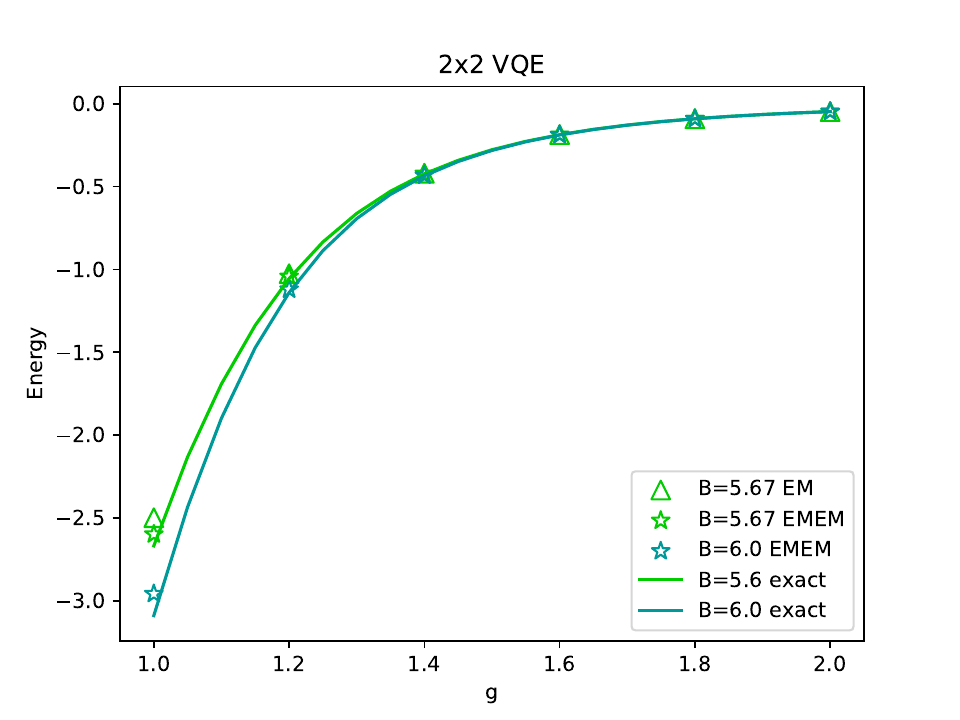}
        \caption{}
        \label{fig:vqe-ground-energy-2x2}
    \end{subfigure}
    \begin{subfigure}{0.48\textwidth}
        \includegraphics[width=\linewidth]{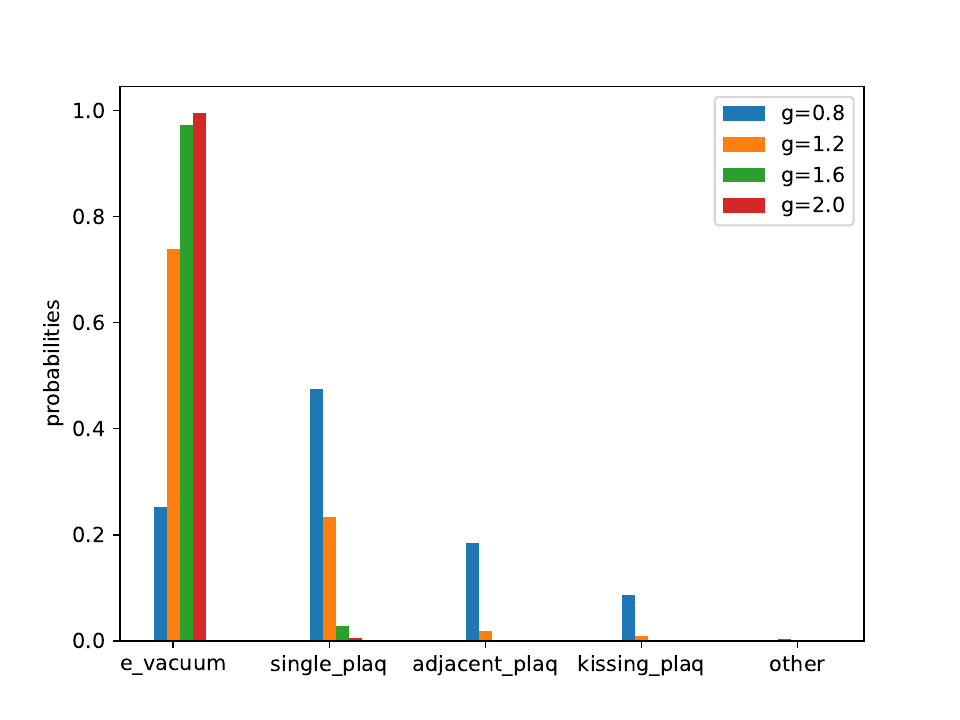}
        \caption{}
        \label{fig:2x2characterization}        
    \end{subfigure}
    \caption{(a) Comparison of VQE with the EM and EMEM ans{\"a}tze with the exact diagonalization of the $2\times2$ plaquettes for $B=5.67$ and $B=6.0$, and (b) Characterization of the states obtained in the $B=6.0$ VQE state preparation using the EMEM ansatz. ``e\_vacuum" refers to the electric vacuum, ``single\_plaq" refers to single plaquette excitations, and ``adjacent\_plaq" and ``kissing\_plaq" refer to the two type of double plaquette excitations possible in the $2\times2$ lattice, which are the excitations that share an edge and a vertex respectively.}
    \label{fig:2x2_vqe}
\end{figure}

As an example of variational methods in two dimensions, we consider the $2\times 2$ lattice with periodic boundary conditions. In Fig.~\ref{fig:vqe-ground-energy-2x2} we compare the energy obtained by exact diagonalization for different values of the $B$ cutoff to variational fits of the ans\"{a}tze EM and EMEM. Since periodic boundary conditions introduce a repeated control link structure in the $2 \times 2$ lattice which is cumbersome to examine perturbatively-- similar to the $2 \times 1$ case-- we restrict the VQE analysis to exclude the multi-Givens ansatz. Furthermore, the adiabatic evolution methods detailed in Sec.~\ref{sec:cube-results} were not applied to the $2 \times 2$ lattice due to the huge circuit depth of the time evolution circuits ($\sim 10^6$ two-qubit gates per trotter step). Nevertheless, the EM and EMEM ans\"{a}tze demonstrate sufficient efficacy in recovering ground state energies for moderate $B$ cutoffs, as illustrated in Fig.~\ref{fig:vqe-ground-energy-2x2}.

To get a sense for the makeup of the ground state, Fig.~\ref{fig:2x2characterization} shows the probabilities of finding different excited states in the variational ground state  with $B=6$. While the probability of finding single plaquette and double plaquette excitations in the strong-coupling limit $g=2$ is vanishingly small, the probabilities to find these excitations steadily increase as one approaches $g\sim1$; in fact, the single plaquette excitations dominate the electric vacuum in the ground state for $g=0.8$. However, the double-plaquette excitations are subdominant even for $g=0.8$, explaining the surprising performance of the EMEM ansatz towards capturing the ground state for weak couplings. 

\subsection{5-plaquette chain}
\begin{figure}[!ht]
    \centering
    \includegraphics[width=0.6\linewidth]{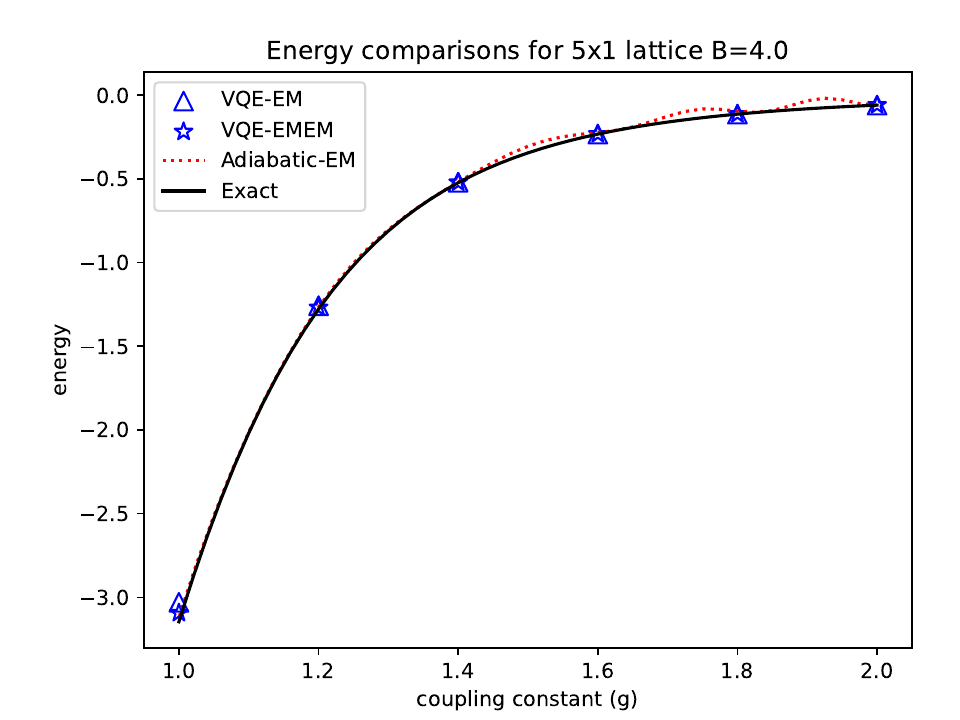}
    \caption{Comparisons of ground state energies computed with the variational EM and EMEM methods, exact diagonalization, and hybrid adiabatic state preparation,  for the 5-plaquette lattice with $B=4.0$ truncation.}
    \label{fig:5x1energy}
\end{figure}

For $N$-plaquette chains, with fixed  classical memory resources, there is a tradeoff between the maximum $N$ and $B$ that may be simulated, i.e., as $N$ increases the maximum $B$ that can be simulated decreases. Fig.~\ref{fig:5x1energy} illustrates the ground state energy for a $5 \times 1$ plaquette chain with $B=4$, comparing exact values against results from adiabatic state preparation and variational optimization of EM and EMEM ansätze. While the multi-Givens ansatz is theoretically well-suited for larger chains due to its shallow circuit depth and scalability, it was not utilized for the $5 \times 1$ case in this study. We intend to investigate its performance on extended plaquette chains in future work.

With a global encoding of the Hilbert space, instead of a local encoding, and a low irrep truncation, it is possible to study chains of $10+$ plaquettes. These analyses, along with a more extensive comparison of global vs. local approaches, will be presented elsewhere. 

\subsection{Quantum resource cost summary}
 We conclude this section with Table~\ref{tab:resource_costs}, sketching the approximate quantum resource costs of preparing ground states for the cube, $2\times 2$, and 5 plaquette lattices using the methods described above. 

 The resources needed  to prepare the ground state generally depend on (1) the number of qubits needed to encode the lattice, and (2) the circuit depth of one Trotter step, which directly controls the circuit depth of the VQE and adiabatic circuits. One Trotter step for the cube with open boundary conditions has a CX gate count $\sim 26{,}000$, which also corresponds to the gate count of the EM preparation circuits (denoted by VQE-EM) because applying the EM operator is equivalent to one Trotter step. The gate cost of one Trotter step can be reduced to about $\sim 8{,}000$ by choosing the Givens rotations that are ``important" and individually parameterizing the rotations (this method is denoted by VQE-MTI). The EMEM preparation circuits have twice the gate count of one Trotter step. The adiabatic preparation circuits, however, are significantly deeper than the variational circuits, because they correspond to many Trotter steps. For example, the adiabatic evolution from $g=4.0$ to $g=1.8$ with adiabatic step size $\Delta g = 0.01$ requires 320 steps, thereby increasing the depth of the adiabatic preparation circuits by a factor of $\sim$ 300. Choosing a hybrid approach allows us to start from a lower value of $g$ and increase the adiabatic step size to $\Delta g = 0.025$ without significant losses in ground state fidelity, while substantially cutting down  the circuit depth.

 The $2\times2$ lattice with periodic boundary conditions has a large number of matrix elements at energy truncations higher than $T_1$, as reflected in Fig.~\ref{fig:unsym_counts}: there are $\sim 10^5$ matrix elements for the $B=5.67$ and $B=6.0$ truncations. This translates to gate counts of  order  a million for the EM and EMEM circuits for this lattice. Even though the gate counts for $B=5.67$ and $B=6.0$ are similar, performing a VQE for $B=6.0$ proves to be computationally harder because encoding a $B=6.0$ lattice requires 28 qubits (3 qubits per link for the irreps $\{1, 3, \bar{3}, 8, 6,\bar{6} \}$ and 4 site qubits)  compared to  20 qubits for $B=5.67$ (2 qubits per link for the irreps $\{1,3,\bar{3},8\}$ and 4 site qubits). It should also be possible to substantially reduce the gate cost associated with the large number of magnetic transitions; the simplest method is to cut out transitions with small matrix elements.

 The $5\times1$ lattice with the $B=4.0$ truncation has a large qubit cost:~30 qubits are needed to encode the 15 links with periodic boundary conditions. However, the number of matrix elements in $N$-plaquette chains with a low energy truncation is significantly lower than the two dimensional lattice because of the simplicity of the site singlets. For $B=4.0$ each site in the $N$ plaquette chain has a maximum multiplicity of 1. This, combined with a straightforward control scheme for the plaquette chain lattices, leads to Trotter step gate counts $\sim 10^4$ for the 5 plaquette chain, 100 times smaller than the $2\times2$ case. The N-plaquette chain therefore seems to be the ideal candidate to test the ground state preparation methods developed in earlier sections for larger lattice sizes. However we also note that $N\times1$ lattices are likewise more suitable for classical tensor network methods.

 It becomes imperative to use ans\"{a}tze with more parameters and shallower circuits for the preparation of ground states on bigger lattices with larger energy cutoffs. While perturbative calculations for smaller lattices with periodic boundary conditions are unwieldy, the perturbation theory informed multi-Givens ansatz is the best candidate for preparing ground states for larger lattices in a scalable manner. In this work, we illustrated the multi-Givens ansatz with the cubic lattice with open boundary conditions; we defer comprehensive benchmarking analysis on bigger systems to future work. 
\begin{table}[!ht]
    \centering
    \begin{tabular}{|c|c|c|c|}
        \hline
        Lattice & Qubits & Method & CX Gate Count   \\
         \hline \hline
         \multirow{4}{*}{Cube ($B=4.0$) } & \multirow{4}{*}{24} & VQE-MTI & 8,000  \\
         \cline{3-4}
         & & VQE-EM & 26,000  \\
         \cline{3-4}
         & & Adiabatic & $8 \times 10^6$ \\
         \cline{3-4}
         & & Hyb-Ad & $6 \times 10^5 $   \\
         \hline \hline 
         \multirow{2}{*}{$2\times2$ ($B=5.67$)} & \multirow{2}{*}{20} & VQE-EM & $1 \times 10^6$     \\
         \cline{3-4}
         & & VQE-EMEM & $2 \times 10^6$\\ 
         \hline \hline
         \multirow{2}{*}{$2\times2$ ($B=6.0$)} & \multirow{2}{*}{28} & VQE-EM &  $1.8\times 10^6$    \\
         \cline{3-4}
         & & VQE-EMEM & $3.6\times10^6$  \\
         \hline \hline
         \multirow{3}{*}{$5\times1$ ($B=4.0$)} & \multirow{3}{*}{30} & VQE-EM &  $18{,}000$    \\
         \cline{3-4}
         & & VQE-EMEM & $36{,}000$ \\
         \cline{3-4}
         & & Hyb-Ad &  $7 \times 10^5$ \\
         \hline
    \end{tabular}
    \caption{Quantum resource costs for different methods of state preparation.}
    \label{tab:resource_costs}
\end{table}

\section{Simulation code: \texttt{ymcirc}}
\label{sec:ymcirc}
We conclude by describing \verb|ymcirc|, a Python code package for $SU(3)$ quantum simulations which we make available with this work. \verb|ymcirc| includes tools for automated construction of time-evolution and ground-state preparation circuits, as well as the interpretation of circuit measurement data. At the time of writing, the latest active development release may be accessed at \url{https://github.com/hepqis-uiuc/ymcirc}.\footnote{Magnetic Hamiltonian matrix element data present in \texttt{ymcirc} were precomputed using the \texttt{pyclebsch} Python package, which we have also made available at \url{https://github.com/hepqis-uiuc/pyclebsch}.}

The supplementary material for this paper consists of a Jupyter Notebook demonstrating the use of \verb|ymcirc| to generate, simulate, and interpret lattice time-evolution circuits.

\subsection{Available functionality}
One of the primary goals of \verb|ymcirc| is to automate to the greatest degree possible the complicated bit/qubit indexing which arises when constructing lattice simulation circuits or interpreting simulation results. This is achieved via separate classes which (1) perform encoding or decoding between physical link or plaquette irrep data and bit strings, (2) arrange \verb|qiskit|'s quantum register classes into an logical lattice data structure which can be indexed via lattice coordinates, (3) arrange measurements\footnote{Both decoded physical irrep data or the underlying bit string measurement data are included.} of a circuit constructed with \verb|ymcirc| into an analogous lattice data structure indexable by lattice coordinates, (4) functionality for computing electric energy eigenvalues, (5) type aliases which serve as aids to the user for constructing valid irrep weight and multiplicity data, and physical plaquette states, and (6) pre-defined irrep weights for $\mathbf{1},\ \mathbf{3},\ \mathbf{\bar 3},\ \mathbf{8},\ \mathbf{6},\ \text{and}\ \mathbf{\bar 6}$.

Additional functionality in the development version of \verb|ymcirc| available at time of writing includes:
\begin{itemize}
    \item Circuit generation for first-order Trotterization of electric and magnetic time evolution for periodic lattices in $d=3/2$ and $d=2$, with step size and magnetic coupling parameterized per-Trotter step. There is also an option to expose each individual Givens rotation in a magnetic Trotter step as a tuneable parameter for VQE state prep.
    \item Irrep truncation magnetic Hamiltonian data for $T_1$ and $T_2$ in $d=3/2$, and $T_1$ in $d=2$.
    \item Magnetic Hamiltonian matrix element data computed from symmetrized CGCs.
    \item Primitive functions for generating individual Givens rotation circuits.
    \item Classes for defining the geometry of lattices, as well as traversing data on a lattice in a consistent order.
    \item Circuit-depth reduction methods from \cite{BalajiEtal:2025:su3circuits} such as control pruning, control fusion, and gray ordering of gates, as well as an optional `v-chain' construction of $MCX$ gates via ancilla qubits \cite{BalaucaArusoaie:2022:efficient_MCQG, BarencoEtal:1995:elementary_gates}).
\end{itemize}

\subsection{Development roadmap}
At present, the circuit generation capabilities of \verb|ymcirc| are limited to $d=3/2$ and $d=2$ for a small selection of irrep truncation schemes. This informs what is the most immediate future development goal: to extend circuit construction to higher dimensionality, and include a greater variety of irrep truncation options (such as the $B$-parameter scheme used in this paper).

Other items in near-to-intermediate term development pipeline include:
\begin{itemize}
    \item $F$-orderings (needed for $d \geq 3$ circuit generation).
    \item Open/mixed boundary conditions.
    \item Rectangular lattices for $d \geq 2$.
    \item Classes for specifying local circuit measurements via lattice coordinates.
    \item Classes for computing observables from local measurement results.
    \item (Longer-term) Incorporating matter degrees of freedom.
\end{itemize}

\section{Conclusion and Future Directions}

We conclude by summarizing the technical and numerical results reported in this work toward the quantum simulation of $SU(3)$ LGT, and some directions for future work. On the technical front, we utilize and extend the reduced electric basis introduced in \cite{BalajiEtal:2025:su3circuits} by introducing a more fine-grained irrep truncation scheme (the ``$B$'' truncations), and by reducing the number of $\Box$ matrix elements through the permutation group properties of the $SU(3)$ irreps. We also introduce variational  ans{\"a}tze, based on strong coupling perturbation theory, for use in ground state preparation at $g\sim 1$. We benchmark  this framework on plaquette chains, the $2\times2$-plaquette lattice, and the 6-plaquette cube with open boundary conditions, performing noiseless simulations on systems of up to 30 qubits. Variational methods, adiabatic state preparation, and hybrid variational-adiabatic methods are found to achieve fidelities $1-F$ better than a few percent at $g=1$.  Finally, alongside this work, we introduce a public-use python suite, \texttt{ymcirc}, to efficiently generate and run lattice evolution circuits in the reduced electric basis.

There are several directions for extension.  To apply variational methods to larger systems, we need a stitching procedure, as in~\cite{Ciavarella:2022:su3_state_prep}, and/or an efficient and systematic way to measure the energy expectation value of the system (or stitchable subsystems), $\text{Tr}\left(H \, \!\rho(\theta)\!\right)$, on a quantum device, where $\rho$ is an output VQE state at each iteration. There are various methods to measure the expectation value, including random basis sampling and quantum state tomography \cite{KohdaEtal:2022:comp_basis_sampling, cramer2010efficient}, but  generally they incur a high cost in state preparation circuits. It remains to be seen whether such methods can be tailored in an efficient way for LGT simulations. For larger systems it may also be useful to work at higher orders in perturbation theory when formulating variational ansatz states; this extension should be straightforward.

It may also be useful to examine hybrid variational-adiabatic approaches to ground state preparation more deeply~\cite{chen2020demonstration, harwood2022improving}. It would be valuable, for example, to have a procedure to estimate the optimal switching point between variational and adiabatic methods. A more efficient hybrid method might also be imagined, in which layers of variational ansatz circuits are systematically added in order to achieve the desired error threshold at lower values of the coupling constant $g$. With such an approach, it might be possible to retain the advantage of relatively shallow VQE circuits while also  controlling errors. The efficacy of the multi-Givens ansatz also remains to be verified for larger lattices and higher energy truncations.

Finally, a substantial challenge in performing interesting simulations on real hardware, or more accurate simulations on future hardware, is the large gate depth associated primarily with the plaquette-local magnetic Hamiltonian. Improving the quantum algorithms used to encode plaquette operators would have a tangible impact on the simulation approach used here. 

\section*{Acknowledgments}
We acknowledge the support of the U.S. Department of Energy, Office of Science, Office of High Energy Physics Quantum Information Science Enabled Discovery (QuantISED) program. This work used the Delta system at the National Center for Supercomputing Applications through allocation PHY230137 from the Advanced Cyberinfrastructure Coordination Ecosystem: Services \& Support (ACCESS) program, which is supported by National Science Foundation grants \#2138259, \#2138286, \#2138307, \#2137603, and \#2138296.

\bibliographystyle{utphys}
\bibliography{refs}

@article{k9p6-c649,
  title = {Efficient Finite-Resource Formulation of Non-Abelian Lattice Gauge Theories beyond One Dimension},
  author = {Fontana, Pierpaolo and Miranda-Riaza, Marc and Celi, Alessio},
  journal = {Phys. Rev. X},
  volume = {15},
  issue = {3},
  pages = {031065},
  numpages = {29},
  year = {2025},
  month = {Sep},
  publisher = {American Physical Society},
  doi = {10.1103/k9p6-c649},
  url = {https://link.aps.org/doi/10.1103/k9p6-c649}
}

@article{Kirby2023exactefficient,
  doi = {10.22331/q-2023-05-23-1018},
  url = {https://doi.org/10.22331/q-2023-05-23-1018},
  title = {Exact and efficient {L}anczos method on a quantum computer},
  author = {Kirby, William and Motta, Mario and Mezzacapo, Antonio},
  journal = {{Quantum}},
  issn = {2521-327X},
  publisher = {{Verein zur F{\"{o}}rderung des Open Access Publizierens in den Quantenwissenschaften}},
  volume = {7},
  pages = {1018},
  month = may,
  year = {2023}
}

@article{Klco:2019evd,
    author = "Klco, Natalie and Stryker, Jesse R. and Savage, Martin J.",
    title = "{SU(2) non-Abelian gauge field theory in one dimension on digital quantum computers}",
    eprint = "1908.06935",
    archivePrefix = "arXiv",
    primaryClass = "quant-ph",
    reportNumber = "INT-PUB-19-033",
    doi = "10.1103/PhysRevD.101.074512",
    journal = "Phys. Rev. D",
    volume = "101",
    number = "7",
    pages = "074512",
    year = "2020"
}

@article{Banuls:2017ena,
    author = {Ba{\~n}uls, Mari Carmen and Cichy, Krzysztof and Cirac, J. Ignacio and Jansen, Karl and K{\"u}hn, Stefan},
    title = "{Efficient basis formulation for 1+1 dimensional SU(2) lattice gauge theory: Spectral calculations with matrix product states}",
    eprint = "1707.06434",
    archivePrefix = "arXiv",
    primaryClass = "hep-lat",
    reportNumber = "DESY-17-108",
    doi = "10.1103/PhysRevX.7.041046",
    journal = "Phys. Rev. X",
    volume = "7",
    number = "4",
    pages = "041046",
    year = "2017"
}

@article{BauerEtal:2023:QS4HEP,
  title = {Quantum Simulation for High-Energy Physics},
  author = {Bauer, Christian W. and Davoudi, Zohreh and Balantekin, A. Baha and Bhattacharya, Tanmoy and Carena, Marcela and de Jong, Wibe A. and Draper, Patrick and El-Khadra, Aida and Gemelke, Nate and Hanada, Masanori and Kharzeev, Dmitri and Lamm, Henry and Li, Ying-Ying and Liu, Junyu and Lukin, Mikhail and Meurice, Yannick and Monroe, Christopher and Nachman, Benjamin and Pagano, Guido and Preskill, John and Rinaldi, Enrico and Roggero, Alessandro and Santiago, David I. and Savage, Martin J. and Siddiqi, Irfan and Siopsis, George and Van Zanten, David and Wiebe, Nathan and Yamauchi, Yukari and Yeter-Aydeniz, K\"ubra and Zorzetti, Silvia},
  journal = {PRX Quantum},
  volume = {4},
  issue = {2},
  pages = {027001},
  numpages = {70},
  year = {2023},
  month = {May},
  publisher = {American Physical Society},
  doi = {10.1103/PRXQuantum.4.027001},
  url = {https://link.aps.org/doi/10.1103/PRXQuantum.4.027001}
}

@article{BauerEtal:2023:QS_fund_particles_forces,
   title={Quantum simulation of fundamental particles and forces},
   volume={5},
   ISSN={2522-5820},
   url={http://dx.doi.org/10.1038/s42254-023-00599-8},
   DOI={10.1038/s42254-023-00599-8},
   number={7},
   journal={Nature Reviews Physics},
   publisher={Springer Science and Business Media LLC},
   author={Bauer, Christian W. and Davoudi, Zohreh and Klco, Natalie and Savage, Martin J.},
   year={2023},
   month=jun,
   pages={420–432}
}

@article{DiMeglioEtal:2024:QC4HEP,
  title = {Quantum Computing for High-Energy Physics: State of the Art and Challenges},
  author = {Di Meglio, Alberto and Jansen, Karl and Tavernelli, Ivano and Alexandrou, Constantia and Arunachalam, Srinivasan and Bauer, Christian W. and Borras, Kerstin and Carrazza, Stefano and Crippa, Arianna and Croft, Vincent and de Putter, Roland and Delgado, Andrea and Dunjko, Vedran and Egger, Daniel J. and Fern\'andez-Combarro, Elias and Fuchs, Elina and Funcke, Lena and Gonz\'alez-Cuadra, Daniel and Grossi, Michele and Halimeh, Jad C. and Holmes, Zo\"e and K\"uhn, Stefan and Lacroix, Denis and Lewis, Randy and Lucchesi, Donatella and Martinez, Miriam Lucio and Meloni, Federico and Mezzacapo, Antonio and Montangero, Simone and Nagano, Lento and Pascuzzi, Vincent R. and Radescu, Voica and Ortega, Enrique Rico and Roggero, Alessandro and Schuhmacher, Julian and Seixas, Joao and Silvi, Pietro and Spentzouris, Panagiotis and Tacchino, Francesco and Temme, Kristan and Terashi, Koji and Tura, Jordi and T\"uys\"uz, Cenk and Vallecorsa, Sofia and Wiese, Uwe-Jens and Yoo, Shinjae and Zhang, Jinglei},
  journal = {PRX Quantum},
  volume = {5},
  issue = {3},
  pages = {037001},
  numpages = {49},
  year = {2024},
  month = {Aug},
  publisher = {American Physical Society},
  doi = {10.1103/PRXQuantum.5.037001},
  url = {https://link.aps.org/doi/10.1103/PRXQuantum.5.037001}
}

@misc{HalimehEtal:2025:QS_out_of_equilibrium,
      title={Quantum simulation of out-of-equilibrium dynamics in gauge theories}, 
      author={Jad C. Halimeh and Niklas Mueller and Johannes Knolle and Zlatko Papić and Zohreh Davoudi},
      year={2025},
      eprint={2509.03586},
      archivePrefix={arXiv},
      primaryClass={quant-ph},
      url={https://arxiv.org/abs/2509.03586}, 
}

@article{LammLawrenceYamauchi:2020:partonQC,
  title = {Parton physics on a quantum computer},
  author = {Lamm, Henry and Lawrence, Scott and Yamauchi, Yukari},
  collaboration = {NuQS Collaboration},
  journal = {Phys. Rev. Res.},
  volume = {2},
  issue = {1},
  pages = {013272},
  numpages = {7},
  year = {2020},
  month = {Mar},
  publisher = {American Physical Society},
  doi = {10.1103/PhysRevResearch.2.013272},
  url = {https://link.aps.org/doi/10.1103/PhysRevResearch.2.013272}
}

@misc{ChenChenMeher:2025:PDF_on_QC,
      title={Parton Distributions on a Quantum Computer}, 
      author={Jiunn-Wei Chen and Yu-Ting Chen and Ghanashyam Meher},
      year={2025},
      eprint={2506.16829},
      archivePrefix={arXiv},
      primaryClass={hep-lat},
      url={https://arxiv.org/abs/2506.16829}, 
}

@misc{BarataRico:2025:jets_entropy_production,
      title={Real-time simulation of jet energy loss and entropy production in high-energy scattering with matter}, 
      author={João Barata and Enrique Rico},
      year={2025},
      eprint={2502.17558},
      archivePrefix={arXiv},
      primaryClass={hep-ph},
      url={https://arxiv.org/abs/2502.17558}, 
}

@article{FarrellIllaSavage:2025:hadronization,
  title = {Steps toward quantum simulations of hadronization and energy loss in dense matter},
  author = {Farrell, Roland C. and Illa, Marc and Savage, Martin J.},
  journal = {Phys. Rev. C},
  volume = {111},
  issue = {1},
  pages = {015202},
  numpages = {21},
  year = {2025},
  month = {Jan},
  publisher = {American Physical Society},
  doi = {10.1103/PhysRevC.111.015202},
  url = {https://link.aps.org/doi/10.1103/PhysRevC.111.015202}
}

@article{DavoidiMuellerPowers:2023:QCphase_diagrams,
  title = {Towards Quantum Computing Phase Diagrams of Gauge Theories with Thermal Pure Quantum States},
  author = {Davoudi, Zohreh and Mueller, Niklas and Powers, Connor},
  journal = {Phys. Rev. Lett.},
  volume = {131},
  issue = {8},
  pages = {081901},
  numpages = {11},
  year = {2023},
  month = {Aug},
  publisher = {American Physical Society},
  doi = {10.1103/PhysRevLett.131.081901},
  url = {https://link.aps.org/doi/10.1103/PhysRevLett.131.081901}
}

@misc{ThanEtal:2024:1dQCD_phase_diagram,
      title={The phase diagram of quantum chromodynamics in one dimension on a quantum computer}, 
      author={Anton T. Than and Yasar Y. Atas and Abhijit Chakraborty and Jinglei Zhang and Matthew T. Diaz and Kalea Wen and Xingxin Liu and Randy Lewis and Alaina M. Green and Christine A. Muschik and Norbert M. Linke},
      year={2024},
      eprint={2501.00579},
      archivePrefix={arXiv},
      primaryClass={quant-ph},
      url={https://arxiv.org/abs/2501.00579}, 
}

@article{KogutSusskind:1975:KS_hamiltonian,
  title = {Hamiltonian formulation of Wilson's lattice gauge theories},
  author = {Kogut, John and Susskind, Leonard},
  journal = {Phys. Rev. D},
  volume = {11},
  issue = {2},
  pages = {395--408},
  numpages = {0},
  year = {1975},
  month = {Jan},
  publisher = {American Physical Society},
  doi = {10.1103/PhysRevD.11.395},
  url = {https://link.aps.org/doi/10.1103/PhysRevD.11.395}
}

@article{ZoharBurrello:2015:LGT4QS,
  title = {Formulation of lattice gauge theories for quantum simulations},
  author = {Zohar, Erez and Burrello, Michele},
  journal = {Phys. Rev. D},
  volume = {91},
  issue = {5},
  pages = {054506},
  numpages = {15},
  year = {2015},
  month = {Mar},
  publisher = {American Physical Society},
  doi = {10.1103/PhysRevD.91.054506},
  url = {https://link.aps.org/doi/10.1103/PhysRevD.91.054506}
}

@article{ByrnesYamamoto:2006:sim_LGT_on_QC,
  title = {Simulating lattice gauge theories on a quantum computer},
  author = {Byrnes, Tim and Yamamoto, Yoshihisa},
  journal = {Phys. Rev. A},
  volume = {73},
  issue = {2},
  pages = {022328},
  numpages = {16},
  year = {2006},
  month = {Feb},
  publisher = {American Physical Society},
  doi = {10.1103/PhysRevA.73.022328},
  url = {https://link.aps.org/doi/10.1103/PhysRevA.73.022328}
}

@article{CiavarellaKlcoSavage:2021:trailhead,
  title = {Trailhead for quantum simulation of SU(3) Yang-Mills lattice gauge theory in the local multiplet basis},
  author = {Ciavarella, Anthony and Klco, Natalie and Savage, Martin J.},
  journal = {Phys. Rev. D},
  volume = {103},
  issue = {9},
  pages = {094501},
  numpages = {45},
  year = {2021},
  month = {May},
  publisher = {American Physical Society},
  doi = {10.1103/PhysRevD.103.094501},
  url = {https://link.aps.org/doi/10.1103/PhysRevD.103.094501}
}

@article{BalajiEtal:2025:su3circuits,
  eprinttype = {arXiv},
  eprintclass = {hep-lat},
  eprint = {2503.08866},
  month = {Sep},
  numpages = {19},
  pages = {054511},
  issue = {5},
  volume = {112},
  journal = {Phys. Rev. D},
  author = {Balaji, Praveen and Conefrey-Shinozaki, Cian\'an and Draper, Patrick and Elhaderi, Jason K. and Gupta, Drishti and Hidalgo, Luis and Lytle, Andrew and Rinaldi, Enrico},
  title = {Quantum Circuits for {SU}(3) Lattice Gauge Theory},
  year = {2025},
  doi = {10.1103/k8f6-yft8},
  url = {https://link.aps.org/doi/10.1103/k8f6-yft8},
  publisher = {American Physical Society},
  copyright = {Creative Commons Attribution 4.0 International},
}

@article{CiavarellaBauer:2024:large_Nc,
  title = {Quantum Simulation of SU(3) Lattice Yang-Mills Theory at Leading Order in Large-${N}_{c}$ Expansion},
  author = {Ciavarella, Anthony N. and Bauer, Christian W.},
  journal = {Phys. Rev. Lett.},
  volume = {133},
  issue = {11},
  pages = {111901},
  numpages = {12},
  year = {2024},
  month = {Sep},
  publisher = {American Physical Society},
  doi = {10.1103/PhysRevLett.133.111901},
  url = {https://link.aps.org/doi/10.1103/PhysRevLett.133.111901}
}

@misc{CiavarellaBurbanoBauer:2025:efficient_large_Nc,
      title={Efficient Truncations of SU($N_c$) Lattice Gauge Theory for Quantum Simulation}, 
      author={Anthony N. Ciavarella and I. M. Burbano and Christian W. Bauer},
      year={2025},
      eprint={2503.11888},
      archivePrefix={arXiv},
      primaryClass={hep-lat},
      url={https://arxiv.org/abs/2503.11888}, 
}

@article{RaychowdhuryStryker:2020:LSH_su2,
  title = {Loop, string, and hadron dynamics in SU(2) Hamiltonian lattice gauge theories},
  author = {Raychowdhury, Indrakshi and Stryker, Jesse R.},
  journal = {Phys. Rev. D},
  volume = {101},
  issue = {11},
  pages = {114502},
  numpages = {25},
  year = {2020},
  month = {Jun},
  publisher = {American Physical Society},
  doi = {10.1103/PhysRevD.101.114502},
  url = {https://link.aps.org/doi/10.1103/PhysRevD.101.114502}
}

@article{KadamRaychowdhuryStryker:2023:LSH_su3_quarks,
  title = {Loop-string-hadron formulation of an SU(3) gauge theory with dynamical quarks},
  author = {Kadam, Saurabh V. and Raychowdhury, Indrakshi and Stryker, Jesse R.},
  journal = {Phys. Rev. D},
  volume = {107},
  issue = {9},
  pages = {094513},
  numpages = {31},
  year = {2023},
  month = {May},
  publisher = {American Physical Society},
  doi = {10.1103/PhysRevD.107.094513},
  url = {https://link.aps.org/doi/10.1103/PhysRevD.107.094513}
}

@article{KadamEtal:2025:LSH_su3_trivalent,
  title = {Loop-string-hadron approach to SU(3) lattice Yang-Mills theory: Hilbert space of a trivalent vertex},
  author = {Kadam, Saurabh V. and Naskar, Aahiri and Raychowdhury, Indrakshi and Stryker, Jesse R.},
  journal = {Phys. Rev. D},
  volume = {111},
  issue = {7},
  pages = {074516},
  numpages = {26},
  year = {2025},
  month = {Apr},
  publisher = {American Physical Society},
  doi = {10.1103/PhysRevD.111.074516},
  url = {https://link.aps.org/doi/10.1103/PhysRevD.111.074516}
}

@misc{JiangKlcoMatteo:2025:cube_qudits,
      title={Non-Abelian dynamics on a cube: improving quantum compilation through qudit-based simulations}, 
      author={Jacky Jiang and Natalie Klco and Olivia Di Matteo},
      year={2025},
      eprint={2506.10945},
      archivePrefix={arXiv},
      primaryClass={quant-ph},
      url={https://arxiv.org/abs/2506.10945}, 
}

@article{KavakiLewis:2024:square_to_triamond_su2,
   title={From square plaquettes to triamond lattices for SU(2) gauge theory},
   volume={7},
   ISSN={2399-3650},
   url={http://dx.doi.org/10.1038/s42005-024-01697-4},
   DOI={10.1038/s42005-024-01697-4},
   number={1},
   journal={Communications Physics},
   publisher={Springer Science and Business Media LLC},
   author={Kavaki, Ali H. Z. and Lewis, Randy},
   year={2024},
   pages={208},
   month=jun
}

@article{KavakiLewis:2025:false_vac_decay_triamond,
  title = {False vacuum decay in triamond lattice gauge theory},
  author = {Kavaki, Ali H. Z. and Lewis, Randy},
  journal = {Phys. Rev. D},
  volume = {112},
  issue = {1},
  pages = {014502},
  numpages = {14},
  year = {2025},
  month = {Jul},
  publisher = {American Physical Society},
  doi = {10.1103/1km8-3tc3},
  url = {https://link.aps.org/doi/10.1103/1km8-3tc3}
}

@article{KanEtal:2021:theta_term,
  title = {Investigating a $(3+1)\mathrm{D}$ topological $\ensuremath{\theta}$-term in the Hamiltonian formulation of lattice gauge theories for quantum and classical simulations},
  author = {Kan, Angus and Funcke, Lena and K\"uhn, Stefan and Dellantonio, Luca and Zhang, Jinglei and Haase, Jan F. and Muschik, Christine A. and Jansen, Karl},
  journal = {Phys. Rev. D},
  volume = {104},
  issue = {3},
  pages = {034504},
  numpages = {14},
  year = {2021},
  month = {Aug},
  publisher = {American Physical Society},
  doi = {10.1103/PhysRevD.104.034504},
  url = {https://link.aps.org/doi/10.1103/PhysRevD.104.034504}
}

@misc{KanNam:2022:QCD_QED_QC,
      title={Lattice Quantum Chromodynamics and Electrodynamics on a Universal Quantum Computer}, 
      author={Angus Kan and Yunseong Nam},
      year={2022},
      eprint={2107.12769},
      archivePrefix={arXiv},
      primaryClass={quant-ph},
      url={https://arxiv.org/abs/2107.12769}, 
}

@article{LammLawrenceYamauchi:2019:general_methods,
  title = {General methods for digital quantum simulation of gauge theories},
  author = {Lamm, Henry and Lawrence, Scott and Yamauchi, Yukari},
  collaboration = {NuQS Collaboration},
  journal = {Phys. Rev. D},
  volume = {100},
  issue = {3},
  pages = {034518},
  numpages = {14},
  year = {2019},
  month = {Aug},
  publisher = {American Physical Society},
  doi = {10.1103/PhysRevD.100.034518},
  url = {https://link.aps.org/doi/10.1103/PhysRevD.100.034518}
}

@article{CarenaEtal:2022:improved_hamiltonians,
  title = {Improved Hamiltonians for Quantum Simulations of Gauge Theories},
  author = {Carena, Marcela and Lamm, Henry and Li, Ying-Ying and Liu, Wanqiang},
  journal = {Phys. Rev. Lett.},
  volume = {129},
  issue = {5},
  pages = {051601},
  numpages = {7},
  year = {2022},
  month = {Jul},
  publisher = {American Physical Society},
  doi = {10.1103/PhysRevLett.129.051601},
  url = {https://link.aps.org/doi/10.1103/PhysRevLett.129.051601}
}

@article{GustafsonLammLovelace:2024:primitive_gates_su2bo,
  title = {Primitive quantum gates for an $SU(2)$ discrete subgroup: Binary octahedral},
  author = {Gustafson, Erik J. and Lamm, Henry and Lovelace, Felicity},
  journal = {Phys. Rev. D},
  volume = {109},
  issue = {5},
  pages = {054503},
  numpages = {14},
  year = {2024},
  month = {Mar},
  publisher = {American Physical Society},
  doi = {10.1103/PhysRevD.109.054503},
  url = {https://link.aps.org/doi/10.1103/PhysRevD.109.054503}
}

@article{GustafsonEtal:2024:primitive_gates_su3S108,
  title = {Primitive quantum gates for an $SU(3)$ discrete subgroup: $\mathrm{\ensuremath{\Sigma}}(36\ifmmode\times\else\texttimes\fi{}3)$},
  author = {Gustafson, Erik J. and Ji, Yao and Lamm, Henry and Murairi, Edison M. and Perez, Sebastian Osorio and Zhu, Shuchen},
  journal = {Phys. Rev. D},
  volume = {110},
  issue = {3},
  pages = {034515},
  numpages = {18},
  year = {2024},
  month = {Aug},
  publisher = {American Physical Society},
  doi = {10.1103/PhysRevD.110.034515},
  url = {https://link.aps.org/doi/10.1103/PhysRevD.110.034515}
}

@article{HaaseEtal:2021:resource_efficient,
  doi = {10.22331/q-2021-02-04-393},
  url = {https://doi.org/10.22331/q-2021-02-04-393},
  title = {A resource efficient approach for quantum and classical simulations of gauge theories in particle physics},
  author = {Haase, Jan F. and Dellantonio, Luca and Celi, Alessio and Paulson, Danny and Kan, Angus and Jansen, Karl and Muschik, Christine A.},
  journal = {{Quantum}},
  issn = {2521-327X},
  publisher = {{Verein zur F{\"{o}}rderung des Open Access Publizierens in den Quantenwissenschaften}},
  volume = {5},
  pages = {393},
  month = feb,
  year = {2021}
}

@article{DAndreaEtal:2024:new_basis_su2,
  title = {New basis for Hamiltonian SU(2) simulations},
  author = {D'Andrea, Irian and Bauer, Christian W. and Grabowska, Dorota M. and Freytsis, Marat},
  journal = {Phys. Rev. D},
  volume = {109},
  issue = {7},
  pages = {074501},
  numpages = {32},
  year = {2024},
  month = {Apr},
  publisher = {American Physical Society},
  doi = {10.1103/PhysRevD.109.074501},
  url = {https://link.aps.org/doi/10.1103/PhysRevD.109.074501}
}

@article{GrabowskaKaneBauer:2025:fully_gauge_fixed_su2,
  title = {Fully gauge-fixed SU(2) Hamiltonian for quantum simulations},
  author = {Grabowska, Dorota M. and Kane, Christopher F. and Bauer, Christian W.},
  journal = {Phys. Rev. D},
  volume = {111},
  issue = {11},
  pages = {114516},
  numpages = {35},
  year = {2025},
  month = {Jun},
  publisher = {American Physical Society},
  doi = {10.1103/PhysRevD.111.114516},
  url = {https://link.aps.org/doi/10.1103/PhysRevD.111.114516}
}

@misc{BurbanoBauer:2024:LSH_on_graphs,
      title={Gauge Loop-String-Hadron Formulation on General Graphs and Applications to Fully Gauge Fixed Hamiltonian Lattice Gauge Theory}, 
      author={I. M. Burbano and Christian W. Bauer},
      year={2024},
      eprint={2409.13812},
      archivePrefix={arXiv},
      primaryClass={hep-lat},
      url={https://arxiv.org/abs/2409.13812}, 
}

@article{JordanLeePreskill:2012:QA4QFT,
   title={Quantum Algorithms for Quantum Field Theories},
   volume={336},
   ISSN={1095-9203},
   url={http://dx.doi.org/10.1126/science.1217069},
   DOI={10.1126/science.1217069},
   number={6085},
   journal={Science},
   publisher={American Association for the Advancement of Science (AAAS)},
   author={Jordan, Stephen P. and Lee, Keith S. M. and Preskill, John},
   year={2012},
   month=jun,
   pages={1130–1133}
}

@article{PeruzzoEtal:2014:VQE_original,
   title={A variational eigenvalue solver on a photonic quantum processor},
   volume={5},
   ISSN={2041-1723},
   url={http://dx.doi.org/10.1038/ncomms5213},
   DOI={10.1038/ncomms5213},
   number={1},
   journal={Nature Communications},
   publisher={Springer Science and Business Media LLC},
   author={Peruzzo, Alberto and McClean, Jarrod and Shadbolt, Peter and Yung, Man-Hong and Zhou, Xiao-Qi and Love, Peter J. and Aspuru-Guzik, Alán and O’Brien, Jeremy L.},
   year={2014},
   pages={4213},
   month=jul
}

@article{TillyEtal:2022:VQE_review,
   title={The Variational Quantum Eigensolver: A review of methods and best practices},
   volume={986},
   ISSN={0370-1573},
   url={http://dx.doi.org/10.1016/j.physrep.2022.08.003},
   DOI={10.1016/j.physrep.2022.08.003},
   journal={Physics Reports},
   publisher={Elsevier BV},
   author={Tilly, Jules and Chen, Hongxiang and Cao, Shuxiang and Picozzi, Dario and Setia, Kanav and Li, Ying and Grant, Edward and Wossnig, Leonard and Rungger, Ivan and Booth, George H. and Tennyson, Jonathan},
   year={2022},
   month=nov,
   pages={1–128}
}

@article{AtasEtal:2021:su2_hadrons_vqe,
   title={SU(2) hadrons on a quantum computer via a variational approach},
   volume={12},
   ISSN={2041-1723},
   url={http://dx.doi.org/10.1038/s41467-021-26825-4},
   DOI={10.1038/s41467-021-26825-4},
   number={1},
   journal={Nature Communications},
   publisher={Springer Science and Business Media LLC},
   author={Atas, Yasar Y. and Zhang, Jinglei and Lewis, Randy and Jahanpour, Amin and Haase, Jan F. and Muschik, Christine A.},
   year={2021},
   pages={6499},
   month=nov
}

@article{FarrellEtal:2023:preparations_1dQCD_axial,
  title = {Preparations for quantum simulations of quantum chromodynamics in $1+1$ dimensions. I. Axial gauge},
  author = {Farrell, Roland C. and Chernyshev, Ivan A. and Powell, Sarah J. M. and Zemlevskiy, Nikita A. and Illa, Marc and Savage, Martin J.},
  journal = {Phys. Rev. D},
  volume = {107},
  issue = {5},
  pages = {054512},
  numpages = {43},
  year = {2023},
  month = {Mar},
  publisher = {American Physical Society},
  doi = {10.1103/PhysRevD.107.054512},
  url = {https://link.aps.org/doi/10.1103/PhysRevD.107.054512}
}

@article{ZhangEtal:2023:vqe_superconducting,
  doi = {10.22331/q-2023-10-23-1148},
  url = {https://doi.org/10.22331/q-2023-10-23-1148},
  title = {Simulating gauge theories with variational quantum eigensolvers in superconducting microwave cavities},
  author = {Zhang, Jinglei and Ferguson, Ryan and K{\"{u}}hn, Stefan and Haase, Jan F. and Wilson, C.M. and Jansen, Karl and Muschik, Christine A.},
  journal = {{Quantum}},
  issn = {2521-327X},
  publisher = {{Verein zur F{\"{o}}rderung des Open Access Publizierens in den Quantenwissenschaften}},
  volume = {7},
  pages = {1148},
  month = oct,
  year = {2023}
}

@article{SchusterEtal:2024:schwinger_chemical_potential,
  title = {Studying the phase diagram of the three-flavor Schwinger model in the presence of a chemical potential with measurement- and gate-based quantum computing},
  author = {Schuster, Stephan and K\"uhn, Stefan and Funcke, Lena and Hartung, Tobias and Pleinert, Marc-Oliver and von Zanthier, Joachim and Jansen, Karl},
  journal = {Phys. Rev. D},
  volume = {109},
  issue = {11},
  pages = {114508},
  numpages = {21},
  year = {2024},
  month = {Jun},
  publisher = {American Physical Society},
  doi = {10.1103/PhysRevD.109.114508},
  url = {https://link.aps.org/doi/10.1103/PhysRevD.109.114508}
}

@article{DavoudiHsiehKadam:2024:scattering_wave_packets_hadrons,
  doi = {10.22331/q-2024-11-11-1520},
  url = {https://doi.org/10.22331/q-2024-11-11-1520},
  title = {Scattering wave packets of hadrons in gauge theories: {P}reparation on a quantum computer},
  author = {Davoudi, Zohreh and Hsieh, Chung-Chun and Kadam, Saurabh V.},
  journal = {{Quantum}},
  issn = {2521-327X},
  publisher = {{Verein zur F{\"{o}}rderung des Open Access Publizierens in den Quantenwissenschaften}},
  volume = {8},
  pages = {1520},
  month = nov,
  year = {2024}
}

@misc{CrippaJansenRinaldi:2024:confinement_2dQED,
      title={Analysis of the confinement string in (2 + 1)-dimensional Quantum Electrodynamics with a trapped-ion quantum computer}, 
      author={Arianna Crippa and Karl Jansen and Enrico Rinaldi},
      year={2024},
      eprint={2411.05628},
      archivePrefix={arXiv},
      primaryClass={hep-lat},
      url={https://arxiv.org/abs/2411.05628}, 
}

@article{Ciavarella:2022:su3_state_prep,
  title = {Preparation of the SU(3) lattice Yang-Mills vacuum with variational quantum methods},
  author = {Ciavarella, Anthony N. and Chernyshev, Ivan A.},
  journal = {Phys. Rev. D},
  volume = {105},
  issue = {7},
  pages = {074504},
  numpages = {19},
  year = {2022},
  month = {Apr},
  publisher = {American Physical Society},
  doi = {10.1103/PhysRevD.105.074504},
  url = {https://link.aps.org/doi/10.1103/PhysRevD.105.074504}
}

@article{GrimsleyEtal:2019:ADAPT_VQE,
    author = "Grimsley, Harper R. and Economou, Sophia E. and Barnes, Edwin and Mayhall, Nicholas J.",
    title = "{An adaptive variational algorithm for exact molecular simulations on a quantum computer}",
    eprint = "1812.11173",
    archivePrefix = "arXiv",
    primaryClass = "quant-ph",
    doi = "10.1038/s41467-019-10988-2",
    journal = "Nature Commun.",
    volume = "10",
    pages = "3007",
    year = "2019"
}

@misc{FarrellEtal:2025:scattering_W_states,
      title={Digital quantum simulations of scattering in quantum field theories using W states}, 
      author={Roland C. Farrell and Nikita A. Zemlevskiy and Marc Illa and John Preskill},
      year={2025},
      eprint={2505.03111},
      archivePrefix={arXiv},
      primaryClass={quant-ph},
      url={https://arxiv.org/abs/2505.03111}, 
}

@article{FarrellEtal:2024:scalable_schwinger_100qubits,
    author = "Farrell, Roland C. and Illa, Marc and Ciavarella, Anthony N. and Savage, Martin J.",
    title = "{Scalable Circuits for Preparing Ground States on Digital Quantum Computers: The Schwinger Model Vacuum on 100 Qubits}",
    eprint = "2308.04481",
    archivePrefix = "arXiv",
    primaryClass = "quant-ph",
    reportNumber = "IQuS@UW-21-060, NT@UW-23-13",
    doi = "10.1103/PRXQuantum.5.020315",
    journal = "PRX Quantum",
    volume = "5",
    number = "2",
    pages = "020315",
    year = "2024"
}

@article{FarrellEtal:2024:hadron_schwinger_112qubits,
  title = {Quantum simulations of hadron dynamics in the Schwinger model using 112 qubits},
  author = {Farrell, Roland C. and Illa, Marc and Ciavarella, Anthony N. and Savage, Martin J.},
  journal = {Phys. Rev. D},
  volume = {109},
  issue = {11},
  pages = {114510},
  numpages = {52},
  year = {2024},
  month = {Jun},
  publisher = {American Physical Society},
  doi = {10.1103/PhysRevD.109.114510},
  url = {https://link.aps.org/doi/10.1103/PhysRevD.109.114510}
}

@article{Zemlevskiy:2025:scalable_scalar_120qubits,
  title = {Scalable quantum simulations of scattering in scalar field theory on 120 qubits},
  author = {Zemlevskiy, Nikita A.},
  journal = {Phys. Rev. D},
  volume = {112},
  issue = {3},
  pages = {034502},
  numpages = {49},
  year = {2025},
  month = {Aug},
  publisher = {American Physical Society},
  doi = {10.1103/qr72-51v1},
  url = {https://link.aps.org/doi/10.1103/qr72-51v1}
}

@misc{ChernyshevEtal:2025:pathfinding_neutrinoless_2beta_decay,
      title={Pathfinding Quantum Simulations of Neutrinoless Double-$\beta$ Decay}, 
      author={Ivan A. Chernyshev and Roland C. Farrell and Marc Illa and Martin J. Savage and Andrii Maksymov and Felix Tripier and Miguel Angel Lopez-Ruiz and Andrew Arrasmith and Yvette de Sereville and Aharon Brodutch and Claudio Girotto and Ananth Kaushik and Martin Roetteler},
      year={2025},
      eprint={2506.05757},
      archivePrefix={arXiv},
      primaryClass={quant-ph},
      url={https://arxiv.org/abs/2506.05757}, 
}

@article{GustafsonEtal:2025:surrogate_scalable_vqe_schwinger,
  title = {Surrogate-constructed scalable-circuits adaptive variational quantum eigensolver in the Schwinger model},
  author = {Gustafson, Erik and Sherbert, Kyle and Florio, Adrien and Shirali, Karunya and Chen, Yanzhu and Lamm, Henry and Valgushev, Semeon and Weichselbaum, Andreas and Economou, Sophia E. and Pisarski, Robert D. and Tubman, Norm M.},
  journal = {Phys. Rev. Appl.},
  volume = {23},
  issue = {6},
  pages = {064002},
  numpages = {18},
  year = {2025},
  month = {Jun},
  publisher = {American Physical Society},
  doi = {10.1103/PhysRevApplied.23.064002},
  url = {https://link.aps.org/doi/10.1103/PhysRevApplied.23.064002}
}

@article{AlexEtal:2011:SUN_CGCs,
	author = {Alex, Arne and Kalus, Matthias and Huckleberry, Alan and von Delft, Jan},
	title = "{A numerical algorithm for the explicit calculation of SU({$N$}) and SL({$N,\mathbb{C}$}) Clebsch–Gordan coefficients}",
	journal = {Journal of Mathematical Physics},
	volume = {52},
	number = {2},
	pages = {023507},
	year = {2011},
	month = {02},
	issn = {0022-2488},
	doi = {10.1063/1.3521562},
	url = {https://doi.org/10.1063/1.3521562}
}

@misc{vanLeeuwenCoehnLisser:1992:LIE_manual,
    title = {LiE, A Package for Lie Group Computations},
    author = {M. A. A. van Leeuwen and A. M. Cohen and B. Lisser},
    year = {1992},
    url = {http://www-math.univ-poitiers.fr/~maavl/pdf/LiE-manual.pdf}
}

@article{Fonseca:2021:GroupMath,
   title={GroupMath: A Mathematica package for group theory calculations},
   volume={267},
   ISSN={0010-4655},
   url={http://dx.doi.org/10.1016/j.cpc.2021.108085},
   DOI={10.1016/j.cpc.2021.108085},
   journal={Computer Physics Communications},
   publisher={Elsevier BV},
   author={Fonseca, Renato M.},
   year={2021},
   month=oct,
   pages={108085}
}

@article{AlcockZeilingerWeigert:2017:young_symmetrizers,
    author = {Alcock-Zeilinger, J. and Weigert, H.},
    title = {Compact Hermitian Young projection operators},
    journal = {Journal of Mathematical Physics},
    volume = {58},
    number = {5},
    pages = {051702},
    year = {2017},
    month = {05},
    issn = {0022-2488},
    doi = {10.1063/1.4983478},
    url = {https://doi.org/10.1063/1.4983478}
}

@book{SakuraiNapolitano:2020:QM,
    place={Cambridge},
    edition={3},
    title={Modern Quantum Mechanics},
    publisher={Cambridge University Press},
    author={Sakurai, J. J. and Napolitano, Jim},
    year={2020}
}

@inbook{BalaucaArusoaie:2022:efficient_MCQG,
  title = {Efficient Constructions for Simulating Multi Controlled Quantum Gates},
  year = {2022},
  author = {Balauca, Stefan and Arusoaie, Andreea},
  booktitle = {Computational Science – ICCS 2022},
  publisher = {Springer International Publishing},
  isbn = {9783031087608},
  pages = {179–194},
  doi = {10.1007/978-3-031-08760-8_16},
  url = {http://dx.doi.org/10.1007/978-3-031-08760-8_16},
  ISSN = {1611-3349},
}

@article{BarencoEtal:1995:elementary_gates,
  author = {Barenco, Adriano and Bennett, Charles H. and Cleve, Richard and DiVincenzo, David P. and Margolus, Norman and Shor, Peter and Sleator, Tycho and Smolin, John A. and Weinfurter, Harald},
  title = {Elementary gates for quantum computation},
  journal = {Physical Review A},
  year = {1995},
  volume = {52},
  number = {5},
  month = {nov},
  pages = {3457–3467},
  issn = {1094-1622},
  doi = {10.1103/physreva.52.3457},
  url = {http://dx.doi.org/10.1103/PhysRevA.52.3457},
  publisher = {American Physical Society (APS)},
}

@article{KohdaEtal:2022:comp_basis_sampling,
  title = {Quantum expectation-value estimation by computational basis sampling},
  author = {Kohda, Masaya and Imai, Ryosuke and Kanno, Keita and Mitarai, Kosuke and Mizukami, Wataru and Nakagawa, Yuya O.},
  journal = {Phys. Rev. Res.},
  volume = {4},
  issue = {3},
  pages = {033173},
  numpages = {19},
  year = {2022},
  month = {Sep},
  publisher = {American Physical Society},
  doi = {10.1103/PhysRevResearch.4.033173},
  url = {https://link.aps.org/doi/10.1103/PhysRevResearch.4.033173}
}

@article{cramer2010efficient,
  title={Efficient quantum state tomography},
  author={Cramer, Marcus and Plenio, Martin B and Flammia, Steven T and Somma, Rolando and Gross, David and Bartlett, Stephen D and Landon-Cardinal, Olivier and Poulin, David and Liu, Yi-Kai},
  journal={Nature communications},
  volume={1},
  number={1},
  pages={149},
  year={2010},
  publisher={Nature Publishing Group UK London}
}

@article{chen2020demonstration,
  title={Demonstration of adiabatic variational quantum computing with a superconducting quantum coprocessor},
  author={Chen, Ming-Cheng and Gong, Ming and Xu, Xiaosi and Yuan, Xiao and Wang, Jian-Wen and Wang, Can and Ying, Chong and Lin, Jin and Xu, Yu and Wu, Yulin and others},
  journal={Physical Review Letters},
  volume={125},
  number={18},
  pages={180501},
  year={2020},
  publisher={APS}
}

@article{harwood2022improving,
  title={Improving the variational quantum eigensolver using variational adiabatic quantum computing},
  author={Harwood, Stuart M and Trenev, Dimitar and Stober, Spencer T and Barkoutsos, Panagiotis and Gujarati, Tanvi P and Mostame, Sarah and Greenberg, Donny},
  journal={ACM Transactions on Quantum Computing},
  volume={3},
  number={1},
  pages={1--20},
  year={2022},
  publisher={ACM New York, NY}
}

@book{georgi-2018-lie-algebras-particle-physics,
  author = {Georgi, Howard},
  title = {Lie Algebras in Particle Physics: From Isospin to Unified Theories},
  year = {2018},
  publisher = {CRC Press},
  isbn = {9780429499210},
  doi = {10.1201/9780429499210},
  url = {http://dx.doi.org/10.1201/9780429499210},
  month = {may},
}

@book{fulton-1991-represen,
  author = {Fulton, William},
  address = {New York},
  booktitle = {Representation theory : a first course},
  isbn = {0387974954},
  language = {eng},
  lccn = {90024926},
  publisher = {Springer-Verlag},
  series = {Graduate texts in mathematics ; 129. Readings in mathematics},
  title = {Representation theory: a first course},
  year = {1991},
}

@book{Sengupta2012,
  title = {Representing Finite Groups},
  ISBN = {9781461412311},
  url = {http://dx.doi.org/10.1007/978-1-4614-1231-1},
  DOI = {10.1007/978-1-4614-1231-1},
  publisher = {Springer New York},
  author = {Sengupta,  Ambar N.},
  year = {2012}
}

@article{li-2023-simulatin,
  author = {Li, Andy C. Y. and Macridin, Alexandru and Mrenna, Stephen and Spentzouris, Panagiotis},
  title = {Simulating scalar field theories on quantum computers with limited resources},
  journal = {Physical Review A},
  year = {2023},
  volume = {107},
  pages = {032603},
  numpages = {22},
  number = {3},
  month = {mar},
  issn = {2469-9934},
  doi = {10.1103/physreva.107.032603},
  url = {http://dx.doi.org/10.1103/PhysRevA.107.032603},
  publisher = {American Physical Society (APS)},
}

@misc{bennakhi-2024-analyzin-quantum-circuit,
  author = {Bennakhi, Ahmad and Franzon, Paul and Byrd, Gregory T.},
  title = {Analyzing Quantum Circuit Depth Reduction with Ancilla Qubits in MCX Gates},
  year = {2024},
  doi = {10.48550/ARXIV.2408.01304},
  url = {https://arxiv.org/abs/2408.01304},
  publisher = {arXiv},
  copyright = {Creative Commons Attribution 4.0 International},
}

@book{Nielsen_Chuang_2010, place={Cambridge}, title={Quantum Computation and Quantum Information: 10th Anniversary Edition}, publisher={Cambridge University Press}, author={Nielsen, Michael A. and Chuang, Isaac L.}, year={2010}}

\appendix
\renewcommand\appendixname{APPENDIX}
\renewcommand\thesection{\Alph{section}}
\renewcommand\thesubsection{\arabic{subsection}} 

\setcounter{figure}{0}

\renewcommand{\thefigure}{S\arabic{figure}}

\section{Exact Diagonalization}
\label{appendix:exact_diagonalization}

This appendix gives an overview of simple implementations of exact diagonalization (ED) and how the complexity scales for different lattices and truncations. For Hilbert spaces that are not too large, the Hamiltonian can be exactly diagonalized or exponentiated using common computational linear-algebra libraries. Furthermore the local nature of the electric basis means the matrix representation for the Hamiltonian is sparse and can thus benefit from more memory- and CPU/GPU-efficient algorithms. In this paper we use the \verb|scipy.sparse| package for its versatility and ease-of-use. After picking a lattice dimension and truncation, a matrix representation for the action of the plaquette operator $\square + \square^\dagger$ is built directly from a table of precomputed master formula coefficients in a tensor-product basis of the vertex sites and links. Note that it is often more efficient to encode the state space using the minimal degrees of freedom necessary rather than a direct embedding of the qubitized Hamiltonian. For instance, with a $T_1$ truncation the dimension of the link space is three-dimensional so a ``qutrit" representation is more efficient. The local plaquette operator matrix representation is then embedded in the Hamiltonian of the full Hilbert space with \verb|scipy.sparse.kron| for each distinct plaquette of the system $N_P$. The electric Casimir operator $E^2$ is diagonal and formulaic in this basis and thus the matrix representation is simple to make. Once constructed, the total matrix can be diagonalized using \verb|scipy.sparse.linalg.eigsh| to find the groundstate energy and eigenvector, or exponentiated using \verb|.expm_multiply| to find the dynamical wavefunction $\ket{\psi(t)} = \exp(-i H t)\ket 0$.

\begin{figure}
    \centering
    \includegraphics[width=1\linewidth]{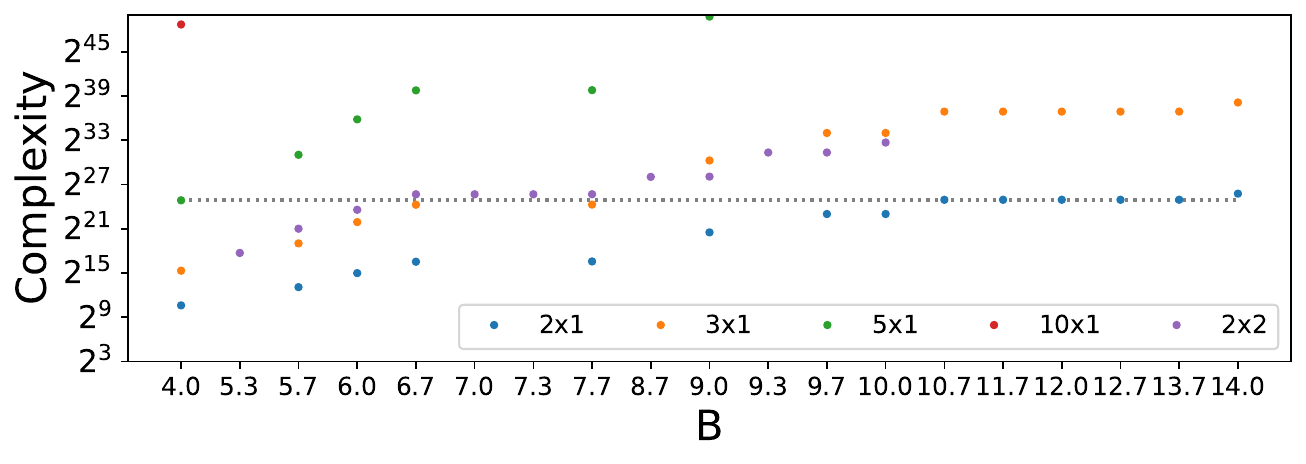}
    \caption{Classical exact diagonalization complexity $\mathcal O(\operatorname{nnz} + n_\text{row})$ for different lattice dimensions e.g. $N\times1$ and truncations at $B$ cuts, assuming the reduced electric basis. The grey line denotes the upper limit of what we could implement on one node of Delta's 128 core CPU with 200GB of memory.}
    \label{fig:ED_complexity}
\end{figure}

These \verb|scipy.sparse| algorithms call matrix multiplication algorithms written in C and C++ which have complexity that scales $\mathcal O(n_\text{nnz} + n_\text{row})$ where $n_\text{nnz}$ is the number of nonzero entries and $n_\text{row}$ is the number of rows in the matrix representation. Because of the large number of non-gauge invariant states in the electric basis and the sparsity of the magnetic term, $n_\text{row}$ is usually of comparable or greater order than $n_\text{nnz}$. For a given energy/irrep truncation given by $B$ (see Section \ref{sec:B-cutoff}), $n_\text{row}(B)$ is $\ell^L\times v^V$ where $\ell$ is the number of irreps, $L$ is the number of links, $v$ is the maximum multiplicity at each vertex, and $V$ is the number of vertices. $n_\text{nnz}(B)$ is $\mathcal O (P\times N_P \times F )+ n_\text{row}$ where $P$ is the number of nonzero entries to the local representation for $\square$, i.e. the number of $\Box$ matrix elements, $N_P$ is the number of plaquettes, and $F$ is the size of the lattice Hilbert space excluding the plaquette. This complexity value is plotted in Fig. \ref{fig:ED_complexity} for different lattices and truncations. Depending on the specific algorithm chosen, the code implementation, the precision required, and the hardware used, resources like memory, CPU/GPU-usage, and wall time will vary. However it is clear that for large lattices with more than the bare minimum gauge-field truncation, the resource requirements quickly exceed the bounds of classical computation.

\section{Strong-Coupling Perturbation Theory Lowest Order Energy}
\label{appendix:strong_PT}
For completeness we provide further details for the perturbation theory computation. The smallest gauge-invariant excitation of the lattice is the excitation of one plaquette $\ket p = (\square + \square^\dagger) \ket 0$ at one site with unperturbed energy eigenvalue
\[E_p = \frac{8g^2}{3a^{d-2}}\]
coming from $\hat H_0 = \frac12 \hat E^2$. We showed in Section \ref{sec:PT} that the normalized perturbed groundstate to lowest nontrivial order in $g$ is
\begin{align*}\ket 0_N &= Z^{-\frac12}\left(\ket{0^{(0)}} + \lambda \ket{0^{(1)}}\right)
\\&= \left(1 + \frac{9N_Pa^{4d-12}}{32g^8}\right)^{-\frac12}\left(\ket{0^{(0)}} + \frac{3 a^{2d-6}}{8g^4} \sum_\text{plaq}\sum_{s=\square,\square^\dagger}\ket {p,s}\right)
\end{align*}
where $N_P$ is the number of distinct plaquettes on the lattice. We can then compute the energy of this state (as a function of $g$) 
\begin{align*}
    E_0(g) &= Z^{-1}\left(\bra{0^{(0)}} + \lambda \bra{0^{(1)}}\right)\left(\hat H_0 + \lambda \hat V\right)\left(\ket{0^{(0)}} + \lambda \ket{0^{(1)}}\right)\\
    &= Z^{-1}\left(\lambda^2 \braket{0^{(1)}}{\hat H_0|0^{(1)}} + 2\lambda^2 \braket{0^{(1)}}{\hat V|0^{(0)}} + \lambda^3\braket{0^{(1)}}{\hat V|0^{(1)}}\right)
\end{align*}
Here we've already removed the $\braket{0^{(0)}}{\hat V|0^{(0)}}$ and $\braket{0^{(0)}}{\hat H_0|0^{(1)}}$ terms due to orthogonality of the excited plaquette states. Looking at each term individually,
\begin{align*}
    \lambda^2\braket{0^{(1)}|H_0}{0^{(1)}} = \frac{\lambda^2}{E_p^2}\sum_{p,s}\sum_{p',s'}\braket{p,s|H_0}{p',s'} = \frac{3N_P a^{3d-10}}{4g^6}
\end{align*}
\begin{align*}
    2\lambda^2\braket{0^{(1)}|V}{0^{(0)}} = -\frac{2\lambda^2}{E_p}\sum_{p,s}\braket{p,s|V}{0^{(0)}} = -\frac{3N_Pa^{3d-10}}{2g^6}
\end{align*}
The last term relies on the fact that $1 = \bra{0^{(0)}}\hat \square^3\ket{0^{(0)}} = \braket{\square^\dagger 0^{(0)}}{\square^2 0^{(0)}}$ which stems from the singlet in $3\otimes 3 \otimes 3$. The $\braket{0^{(1)}}{\hat V|0^{(1)}}$ term can then be interpreted as a sum over double excitations which match a single excitation, or triple excitations that match the ground state.
\begin{align*}
    \lambda^3\braket{0^{(1)}|V}{0^{(1)}} &= \frac{\lambda^3}{E_p^2}\sum_{p,s}\sum_{p',s'}\braket{p,s|V}{p',s'} = \frac{2N_P\lambda^3}{E^2_p} = \frac{9N_P a^{5d-16}}{32g^{10}}
\end{align*}
Thus all together
\begin{align}E_0(g) = -N_P\left(1 + \frac{9N_Pa^{4d-12}}{32g^8}\right)^{-1}\left(\frac{3a^{3d-10}}{4g^6} + \frac{9a^{5d-16}}{32g^{10}}\right)
\end{align}
as written in Eqn~\ref{eq:E0_eqn}.

\section{Permutation symmetries in direct-sum decompositions}
\label{appendix:improved cgcs}

We will briefly review how permutation symmetries emerge when considering direct-sum decompositions as well as the representation theory of the symmetric group $S_n$. Once these preliminaries have been accomplished, we will be able to describe our procedure for computing CGCs in a basis that diagonalizes the permutation symmetries latent in the direct sum decomposition of $SU(N)$ tensor products.

We can understand how permutation symmetries emerge in computing direct-sum decompositions by considering two basic examples. First, there is the familiar example of the antisymmetric singlet and the fully-symmetric triplet appearing in the decomposition of two spin-1/2 particles. Second, consider the decomposition of the tensor product $3\otimes 3$ in $SU(3)$ \cite{georgi-2018-lie-algebras-particle-physics}. The product of two fundamental irreps, $u^i$ and $u^j$, can be written as the sum of a totally symmetric two-index tensor and a totally antisymmetric one,
\begin{equation}
    u^i u^j = \frac{1}{2}\left(u^i u^j + u^j u^i\right) + \frac{1}{2}\epsilon^{ijk}\epsilon_{klm}u^l u^m.
\end{equation}
Symmetric tensors with two fundamental indices have six degrees of freedom, while  antisymmetric tensors have three. The decomposition is:
\begin{equation}
\label{eq:3 times 3 with s_n irreps}
    3\otimes 3 = 6^{(1,1)} \oplus \bar{3}^{(2)},
\end{equation}
where the superscript $\lambda$ on each $SU(3)$ irrep labels the representation of the symmetric group which the $SU(3)$ irrep transforms under. In the current case, $\lambda = (1,1)$ is the fully-symmetric representation of the permutation group $S_2$, while $\lambda = (2)$ is the fully anti-symmetric representation.

In  general, irreps $\lambda$ of the symmetric group $S_n$ are labeled by partitions of $n$, which correspond to Young diagrams. The entries in a partition of $n$ stipulate the number of boxes in each row of the Young diagram. For example,
\begin{equation}
    (4,2,1) \leftrightarrow \ydiagram{4,2,1}
\end{equation}
labels an irreps of $S_7$. The dimensionality of $S_n$ irreps is given by the hook-length formula \cite{fulton-1991-represen}. It is computed via the hook-length $h(i,j)$ of the $(i,j)$-th box in a Young diagram. $h(i,j)$ is the number of boxes in a `hook' connecting all the boxes below the $(i,j)$-th one and all the boxes to the right of it. As an example, we label each box in the $(4,2,1)$ Young diagram with its hook length:
\begin{equation}
    \begin{ytableau}
    6 & 4 & 1 & 1\\
    3 & 1\\
    1
    \end{ytableau}
\end{equation}

The hook-length formula gives the dimensionality of an irrep $\lambda$ of $S_n$ as
\begin{equation}
    \dim \lambda = \frac{n!}{\prod_{(i,j)}h(i,j)}.
\end{equation}
Continuing with our example of $\lambda = (4,2,1)$, we have that
\begin{equation}
    \dim (4,2,1) = \frac{7!}{6\cdot 4 \cdot 1 \cdot 1 \cdot 3 \cdot 1 \cdot 1} = 70.
\end{equation}
By extension it is straightforward to see that the notation appearing in the direct-sum decomposition of $3\otimes 3$ in Eqn.~\ref{eq:3 times 3 with s_n irreps} indicates that the $6$ and $3$ transform in one-dimensional irreps of $S_2$, as they must, since neither irrep is repeated. More generally a term $R^\lambda$ appearing in a direct-sum decomposition denotes an $SU(N)$ irrep $R$ which transforms in a representation $\lambda$ of $S_n$. This notation can be understood as compactly denoting a product of the representations $R$ and $\lambda$ which collectively has dimension $\dim R \cdot \dim \lambda$.\footnote{That the action of $S_n$ and $SU(3)$ on tensor products of $SU(3)$ irreps commutes---and moreover that we can therefore determine a direct-sum decomposition in which each term consists of one $SU(3)$ irrep factor and one $S_n$ irrep factor---is a manifestation of Schur-Weyl duality \cite{Sengupta2012}.}

In the case of $3\otimes 3$, we automatically obtained a direct-sum decomposition whose terms were also irreps of the symmetric group. When considering higher tensor powers of an $SU(N)$ irrep, $R^{\otimes n}$ with $n>2$, this is no longer guaranteed. If we desire direct summands $R'^\lambda$ which are also irreps of the $S_n$, we must explicitly perform some kind of procedure to obtain them.\footnote{In the context where $R'^\lambda$ arises from computing the direct sum decomposition of $R^{\otimes n}$, it is referred to as a \emph{symmetrized tensor product} or \emph{plethysm} \cite{vanLeeuwenCoehnLisser:1992:LIE_manual}.}

We achieve this via a slight modification of the algorithm given in \cite{AlexEtal:2011:SUN_CGCs} for numerically computing $SU(N)$ CGCs. The modification takes place at the point in that algorithm where a highest-weight state of an $SU(N)$ irrep has been obtained from a tensor product of $SU(N)$ irreps. At that point, we construct the set of \emph{Hermitian Young projection operators} described in \cite{AlcockZeilingerWeigert:2017:young_symmetrizers} which project onto irreps of the symmetric group, and apply them all to the highest weight state. This effectively splits the highest weight state into subspaces corresponding to irreps of $S_n$. From this point forward, our computation of CGCs continues to follow the procedure outlined in \cite{AlexEtal:2011:SUN_CGCs}.

As an example of this procedure for `diagonalizing the permutation group symmetry,' consider the direct-sum decomposition of $8\otimes 8\otimes 8$. Without diagonalizing the  permutation group symmetry, we would write
\begin{equation}
\label{eq:triple 8 no perm sym}
    8\otimes 8\otimes 8 = 64 \oplus 35'_2 \oplus 35_2 \oplus 27_6 \oplus 10'_4 \oplus 10_4 \oplus 8_8 \oplus 1_2,
\end{equation}
where subscripts are present to denote nontrivial multiplicities of the irrep, and primes indicate distinct irreps which have the same dimensionality. If we instead use the modified procedure involving Hermitian Young projectors, we find that
\begin{multline}
\label{eq:triple 8 with perm sym}
    8\otimes 8\otimes 8 = 64^{(3)} \oplus 35'^{(2, 1)} \oplus 35^{(2, 1)} \oplus 27^{(1, 1, 1)} \oplus 27^{(2, 1)}_2 \oplus 27^{(3)}\\
    \oplus 10'^{(1, 1, 1)} \oplus 10'^{(2, 1)} \oplus 10'^{(3)} \oplus 10^{(1, 1, 1)} \oplus 10^{(2, 1)} \oplus 10^{(3)}\\
    \oplus 8^{(1, 1, 1)} \oplus 8^{(2, 1)}_3 \oplus 8^{(3)} \oplus 1^{(1,1,1)} \oplus 1^{(3)}.
\end{multline}
We can focus on the subspaces of $8\otimes8\otimes 8$ which transform in the $27$ irrep to illustrate the difference between these results. The direct sum decomposition in Eqn~ \ref{eq:triple 8 no perm sym} has six $27$ irreps, which do not have clear transformation properties under the action of $S_3$ on $8\times 8\times 8$. In contrast, the direct sum decomposition in Eqn.~\ref{eq:triple 8 with perm sym} contains two $27$ $SU(3)$ irreps which transform in one-dimensional irreps of $S_3$ and two pairs of $27$ $SU(3)$ irreps which transform in two-dimensional representations of $S_3$. In both cases, there is a 162-dimensional subspace containing six copies of the $27$ $SU(3)$ irrep, but in the latter case, the CGCs affording the decomposition further block-diagonalize this  subspace according the remaining permutation symmetry.

\end{document}